\pdfoutput=1 

\documentclass[a4paper, 11pt, final]{article}

\usepackage{amssymb} % provides various useful mathematical symbols
\usepackage{amsthm} % provides extended theorem environments
\usepackage{newtxmath} % provides additional math symbol support in Times New Roman
\usepackage{amsmath,empheq}

\DeclareMathAlphabet{\mathcal}{OMS}{cmsy}{m}{n}
\DeclareMathAlphabet\mathbfcal{OMS}{cmsy}{b}{n}

\DeclareFontFamily{U}{dutchcal}{\skewchar\font=45 }
\DeclareFontShape{U}{dutchcal}{m}{n}{<-> s*[1.0] dutchcal-r}{}
\DeclareFontShape{U}{dutchcal}{b}{n}{<-> s*[1.0] dutchcal-b}{}
\DeclareMathAlphabet{\mathcald}{U}{dutchcal}{m}{n}
\SetMathAlphabet{\mathcald}{bold}{U}{dutchcal}{b}{n}
\DeclareMathAlphabet\mathcalz{T1}{pzc}{mb}{it}

\usepackage[utf8]{inputenc}
\usepackage{microtype} % Slightly tweak font spacing for aesthetics
\usepackage{newtxtext} % change font to Adobe Times New Roman
\usepackage[british]{babel}
\usepackage{csquotes}		%Necessary for biber

\providecommand{\JEL}[1]{\textit{\textbf{JEL: }} #1}
\providecommand{\keywords}[1]{\textbf{\textit{Keywords--- }} #1}

\usepackage{setspace} 

\usepackage{parskip} % each new line automatically spaces previously paragraph correctly
\usepackage{etoolbox}
\usepackage{titlesec}
\titleformat{\section}{\normalfont\Large\bfseries}{\thesection}{1em}{}
\titleformat{\subsection}{\normalfont\large\bfseries}{\thesubsection.}{1em}{}
\titleformat{\subsubsection}{\normalfont\normalsize\itshape}{\thesubsubsection.}{1em}{}

\usepackage{abstract}
\renewenvironment{abstract}
 {\normalfont
  \begin{center}
  \bfseries \abstractname\vspace{-.5em}\vspace{0pt}
  \end{center}
  \list{}{
    \setlength{\leftmargin}{0cm}%
    \setlength{\rightmargin}{\leftmargin}%
  }%
  \item\relax}
 {\endlist}

\usepackage{authblk} % package for author affiliations
\usepackage[bottom]{footmisc} % Makes footnotes stick to bottom of the page

\usepackage{graphicx} % More advanced figure inclusion
\usepackage{float} % For specifying table/figure locations, i.e. [ht!]
\usepackage{subcaption}
\usepackage{afterpage} % For encapsulating a floating figure on a single page   

\usepackage{printlen}

\usepackage[labelfont=bf]{caption}
\usepackage{color, colortbl}
\definecolor{LightGray}{rgb}{0.93,0.914,0.914}    
\usepackage{longtable,rotating} % For long tables that span multiple pages
\usepackage{booktabs} % used for making more professional-looking tables
\usepackage{multirow} % used for cell-merging in complex tables
\usepackage{arydshln} % for drawing other line types in tables
\PassOptionsToPackage{hyphens}{url} % for hyphenating URLs in main text

\newcommand*\rot{\rotatebox{90}}

\usepackage{fancyhdr}
\fancyheadoffset{0cm}
\makeatletter
\newcommand{\quickwordcount}[1]{
  \immediate\write18{texcount -quiet -incbib -sub=none -utf8 -1 -sum -merge -encoding=utf8 #1.tex > #1-words}%
  \immediate\openin\somefile=#1-words
  \read\somefile to \@@localdummy
  \immediate\closein\somefile
  \setcounter{wordcounter}{\@@localdummy}
  \@@localdummy
}
\makeatother

\usepackage{silence}
\usepackage[style=apa,backend=biber,natbib,hyperref]{biblatex}
\DeclareLanguageMapping{british}{british-apa}
\defbibenvironment{bibliography}{\enumerate}{\endenumerate}{\item}

\usepackage[colorlinks=false,allcolors=black]{hyperref} 
\let\orgautoref\autoref

\renewcommand{\autoref}[1]
{%
\def\equationautorefname{Eq.}%
\def\figureautorefname{Fig.}%
\def\subfigureautorefname{Fig.}%
\orgautoref{#1}%
}

\usepackage[T1]{fontenc}

\usepackage[nameinlink,capitalise]{cleveref}
\usepackage{algorithm,algpseudocode}

\makeatletter
\newlength{\trianglerightwidth}
\algnewcommand{\LineCommentCont}[1]{\Statex \hskip\ALG@thistlm%
  \parbox[t]{\dimexpr\linewidth-\ALG@thistlm}
{\leftskip=\algorithmicindent
  \hangindent=\algorithmicindent 
  \hangafter=1%
  \strut\makebox[\algorithmicindent][c]{$\triangleright$}#1\strut}
  } % \trianglerightwidth
\makeatother

\usepackage[left=1.8cm, right=1.8cm, bottom=2.5cm, top=2.5cm]{geometry}

\begin{document}

%TC:ignore

% more Hyperref options (must be after \begin{document}
\renewcommand{\figureautorefname}{Fig.}
\onehalfspacing

%--------------------------------------------------------%
%	TITLE PAGE
%--------------------------------------------------------%

%--------------------------------------------------------%
%	TITLE
%--------------------------------------------------------%

% Article title
\newcommand{\MainTitleText}{The loss optimisation of loan recovery decision times using forecast cash flows}

\title{\fontsize{20pt}{0pt}\selectfont\textbf{\MainTitleText
}}

%--------------------------------------------------------%
%	AUTHORS
%--------------------------------------------------------%   

\author[,a]{\large Arno Botha \thanks{ ORC iD: 0000-0002-1708-0153}}
\author[,a]{\large Conrad Beyers \thanks{Corresponding author: \url{conrad.beyers@up.ac.za}}}
\author[b]{\large Pieter de Villiers}
\affil[a]{\footnotesize \textit{Department of Actuarial Science, University of Pretoria, Private Bag X20, Hatfield, 0028, South Africa}}
\affil[b]{\footnotesize \textit{Department of Electrical, Electronic, and Computer Engineering, University of Pretoria, Private Bag X20, Hatfield, 0028, South Africa}}
\renewcommand\Authands{, and }

% Today's date
 	%\date{Submitted: \usvardate\today}
    
%by specifying the below, we essentially "rewrite" the command \maketitle, which is normally called in main.tex.
%this is done primarily to abuse the \date command above

\makeatletter
\renewcommand{\@maketitle}{
    \newpage
     \null
     \vskip 1em%
     \begin{center}%
      {\LARGE \@title \par
      	\@author \par}
     \end{center}%
     \par
 } 
 \makeatother
 
 \maketitle

%--------------------------------------------------------%
%	ABSTRACT
%--------------------------------------------------------%    
{
    \setlength{\parindent}{0cm}
    \rule{1\columnwidth}{0.4pt}
    \begin{abstract}
    A theoretical method is empirically illustrated in finding the best time to forsake a loan such that the overall credit loss is minimised. This is predicated by forecasting the future cash flows of a loan portfolio up to the contractual term, as a remedy to the inherent right-censoring of real-world `incomplete' portfolios. Two techniques, a simple probabilistic model as well as an eight-state Markov chain, are used to forecast these cash flows independently. We train both techniques from different segments within residential mortgage data, provided by a large South African bank, as part of a comparative experimental framework. As a result, the recovery decision's implied timing is empirically illustrated as a multi-period optimisation problem across uncertain cash flows and competing costs. Using a delinquency measure as a central criterion, our procedure helps to find a loss-optimal threshold at which loan recovery should ideally occur for a given portfolio. Furthermore, both the portfolio's historical risk profile and forecasting thereof are shown to influence the timing of the recovery decision. This work can therefore facilitate the revision of relevant bank policies or strategies towards optimising the loan collections process, especially that of secured lending.
    \end{abstract}
     
     % Insert keywords here
     
     % Insert JEL codes here
    \JEL{C44, C53, C61.}
     
    \rule{1\columnwidth}{0.4pt}
}
 
\subsection*{Acknowledgements}
\noindent The authors are grateful towards First National Bank, a division of FirstRand Bank Limited, for providing the data that made this research possible, as well as Andrzej Szanda for facilitating this process. This work is also financially supported by the Absa Chair in Actuarial Science, hosted at the University of Pretoria, with no known conflicts of interest that may have influenced the outcome of this work. The authors would like to thank all anonymous referees for their valuable contributions that improved this work.

\noindent Word count (excluding front matter): 9670 %\quickwordcount{ms} 

\noindent Figure count: 12

%TC:endignore

%--------------------------------------------------------%
%	CONTENT
%--------------------------------------------------------%

\newpage

\section{Introduction}
\label{sec:1}

Loan delinquency is deeply embedded into most of the credit risk modelling exercises of a lender. It is often the broad backbone on which banks construe basic credit and pricing decisions, devise debt collection strategies, and perform overall risk management, in addition to its use within risk modelling. Though it is generally an abstract concept, we define \textit{delinquency} as a time-dependent measurable quantity that represents the extent of eroded trust between bank and borrower. Accordingly, a \textit{delinquency measure} $g$ reflects the degree of non-payment based on the fundamental idea of a borrower owing $I_t>0$ (the instalment) but only repaying $R_t\geq0$ (the receipt) at a specific point in time $t$. When $R_t<I_t$, this measure $g$ should then quantify the extent $I_t-R_t$ of what is essentially the breakdown of trust in honouring the original credit agreement. In practice, the accountancy-based days past due (DPD) is a classical precursor to constructing $g$, whereby the unpaid portion of an instalment is aged into increasingly severe bins as each 30-day calendar month lapses: 30 days, 60 days, 90 days, and so forth. This is more formally defined as  $g_0(t)=f\left(A_t/I\right)$, where $A_t$ is the accumulated arrears amount at time $t$, $I$ is the fixed instalment, and $f$ is a chosen rounding function that maps the input to the number of payments in arrears as the output. This methodical manner of measuring risk using $g$ was especially made necessary to facilitate the advent of automated application scoring during the 1960s. Subsequently, the widespread practice of credit scoring is perhaps the modern bedrock of all mathematical modelling in banking, into which `delinquency' is again embedded.

Accruing delinquency over time will increasingly chafe away at the bank's confidence in the agreement. This erosion continues until reaching a certain threshold $d$, predicated to exist on the domain of $g$, beyond which the obligor is considered as in `default'. That said, these default definitions often contain more qualitative criteria, other than that of simply breaching $d$. Credit risk modelling itself typically uses different definitions given the portfolio type and, more importantly, the modelling context: quantifying either \textit{unexpected} or \textit{expected} credit losses. However, the international standards that govern these contexts, i.e., Basel II and IFRS 9 respectively, are enforced to varying degrees by individual regulators. Accordingly, `default' and the value of $d$ as imposed by the regulator may differ across competing jurisdictions, which certainly complicates any related modelling for multinational banks. Yet even if $d$ is made equal and strictly regulated across all nations, there is little objective evidence for its supposed optimality or suitability. Another challenge presents itself as a consequence of acquiring, merging, and selling-off loan portfolios (or portions thereof) amongst lenders. So-called `legacy' definitions from the previous owners can certainly conflict with that of the new owner, which implies the existence of multiple concurrent definitions in the same portfolio. As a result of all of these difficulties, the very idea of `default' has become an utterly vague and incoherent concept in trying to serve so many `masters' at once. 

However, the original premise of a default definition is that of reaching a so-called "point of no return", beyond which loan repayment becomes extremely doubtful. In principle, this probabilistic point can be unique to a portfolio or bank, given the differences in their risk appetites, market conditions, and the element of time. Once breached, the lender effectively assumes that the obligor's delinquency will perpetuate beyond recovery, should the agreement be retained. Therefore, the bank pursues maximal debt recovery within the shortest time possible, based on the five-phase credit management model of \citet[pp.~11--13, ~147--153]{finlay2010book}. The challenge now becomes to find this ideal switching point or, put differently, the \textit{best} time at which the lender should abandon all hope of repayment and instead pursue debt recovery, including seizing any collateral. Furthermore, finding this point using $g$ is convenient since past loan performance can be projected into scale-invariant delinquency progressions, whilst retaining any behavioural information that may affect the location of the aforementioned point.

\begin{figure}[ht!]
\centering\includegraphics[width=0.6\linewidth,height=0.35\textheight]{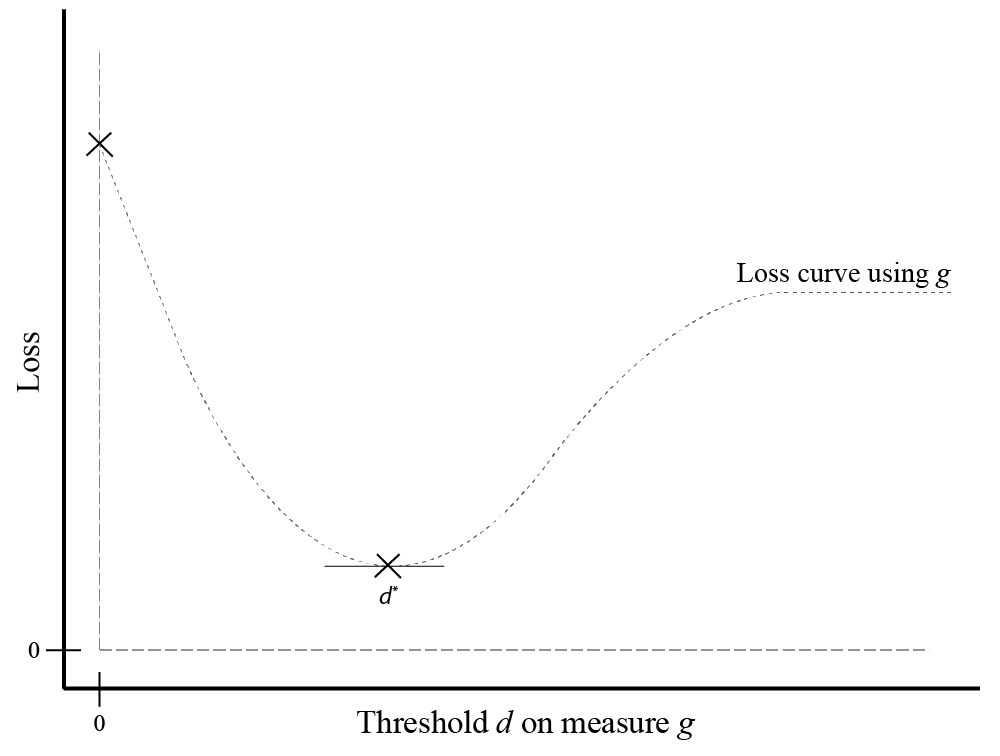}
\caption{Illustrating the LROD-procedure in finding the optimal threshold $d^\ast$ on the domain of a delinquency measure $g$, in optimising the recovery decision's timing.}\label{fig:Loss_Approach}
\end{figure}

In fact, our previous work in \citet{botha2020paper1} explored a more fundamental meaning of `default' as the risk-based "point of no return" beyond which loan collection, if still pursued, becomes financially sub-optimal. The `default' state is simply defined as breaching a variable threshold $d$ on $g$, which deliberately deviates from current practices and relevant regulations. However, this allows for systematically assessing the "net cost" of each candidate threshold, as part of the so-called \textit{Loss-based Recovery Optimisation across Delinquency} (LROD) procedure. Too strict a threshold will marginalise certain accounts that would have resumed payment (or `cured' from delinquency), had the bank not foreclosed (or charged-off) that soon. Too lenient a threshold will naively tolerate increasing amounts in arrears at the cost of higher liquidity risk and greater capital buffers, possibly becoming capital-inefficient. These two extremes are offset against each other using financial loss as basis, thereby forming a proverbial `Goldilocks'-region in which an \textit{ideal} threshold can theoretically exist. The LROD-procedure's output is shown in \autoref{fig:Loss_Approach} as a hypothetical loss curve across candidate thresholds, having attained a global minimum loss at $d^\ast$. In turn, this minimum informs a portfolio's ideal tolerance level across the spectrum of $g$-measurable delinquency before initiating debt recovery, as we demonstrated using simulated toy portfolios.

\begin{figure}[ht!]
\centering\includegraphics[width=0.8\linewidth,height=0.45\textheight]{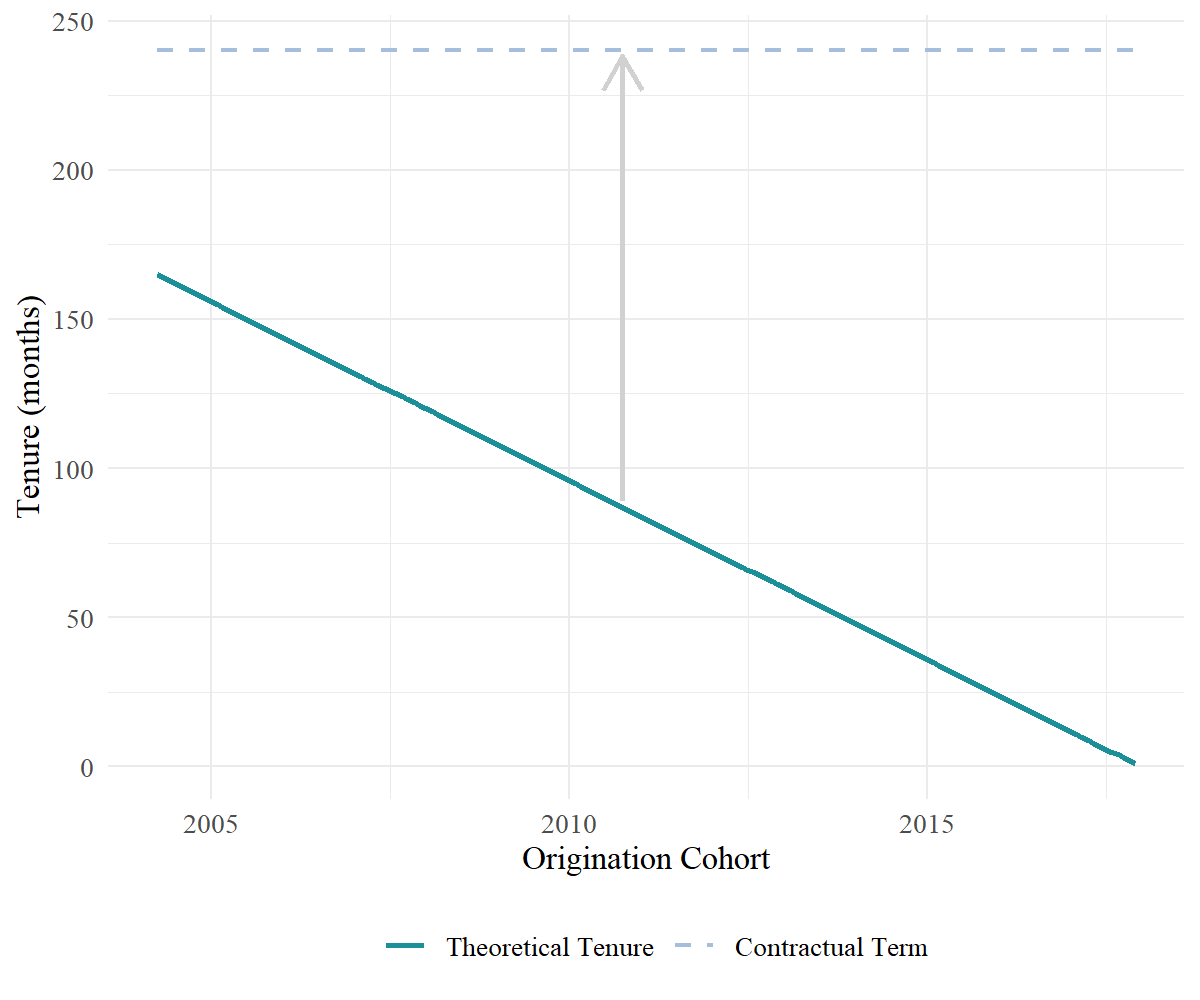}
\caption{Illustrating the increasing right-censoring effect for newer origination cohorts.}\label{fig:MainObjective}
\end{figure}

As our main contribution, we explore the application (and refinement) of the LROD-procedure using a real-world South African mortgage portfolio. This is non-trivial since real-world portfolios are often `incomplete' (or censored) insofar that many constituent loan accounts are not (yet) fully observed up to the contractual term. This is unsurprising since most portfolios are actively being grown every month. However, a truly uncensored portfolio, even if ideal for recovery optimisation, may not reflect current market conditions anymore (especially when considering longer-term loans), which may adversely affect the optimisation. Lastly, the LROD-procedure itself was originally developed within the context of uncensored portfolios. Therefore, as a secondary contribution, we demonstrate a more feasible approach wherein available data is first leveraged to forecast the residual cash flows of each account up to its contractual term, as shown in \autoref{fig:MainObjective}. Performing this necessary step will then enable the empirical use of the LROD-procedure.

Literature is explored in \autoref{sec:2} on default definitions, regulations, and optimising the loan recovery process in general. In turn, this review becomes the basis for framing the bank's recovery decision as an delinquency-based optimisation problem. The LROD-procedure that implements this optimisation problem is briefly discussed in \autoref{sec:3}. We subsequently outline two receipt forecasting techniques as candidate models in \crefrange{sec:3-random_defaults}{sec:3-markovian_defaults}. These candidates include a simple probabilistic technique as well as a more sophisticated eight-state Markov chain. Each technique is parametrised from 20-year residential mortgage data and assessed on its forecasting quality in \autoref{sec:4}. The LROD-procedure is then applied on the now-completed portfolio in \autoref{sec:5}, accompanied by a discussion of the ensuing results. Lastly, we demonstrate a Monte Carlo-based procedural refinement by which the variance of the underlying forecasts can be analysed, thereby granting additional assurance on the stability of any optima. Having concluded in \autoref{sec:6}, the timing of a bank's recovery decision is successfully illustrated as a delinquency-based optimisation problem using real-world data. The source code accompanying this study is published in \citet{botha2020sourcecode2}.
\section{Opportune loan recovery as a delinquency-based optimisation problem}
\label{sec:2}

To collect on a distressed loan is to have breached a certain "point of no return" in the relationship between bank and borrower. This notion is arguably similar to that underlying a default definition, which suggests a valuable starting point for a discussion ultimately centred on \textit{when} to abandon a troubled loan. To this end, various regulations and standards are examined in \autoref{sec:2_regulatory_overview}, followed by surveying the relevant literature in \autoref{sec:2_default_servant_many_masters}.

\subsection{A regulatory overview of default definitions}
\label{sec:2_regulatory_overview}

The Basel II Capital Accords regulate the manner in which banks calculate capital buffers against \textit{unexpected} losses, which is predicated by first quantifying the \textit{expected} loss. Essentially, credit risk is expressed as the product of three particular risk parameters: 1) default risk, or the probability of default (PD); 2) loss risk, or the loss given default (LGD); 3) and exposure risk, or the exposure-at-default (EAD). For a thorough treatise on this topic, refer to \citet[pp.~289--293]{thomas2009book}, \citet[chap.~4,~6]{VanGestel2009book}, and \citet[chap.~5--11]{baesens2016credit}. Clearly, the notion of `default' is again embedded within in all three components. However, the manner in which `default' is defined generally varies by product, customer type, and bank. Some definitions include filing for bankruptcy, unfulfilled claims, negative net present values, becoming and remaining overdrawn for a period, as well as being three payments in arrears, i.e., $d=3$. This last definition, or similarly, 90 DPD, is perhaps most commonly used in amortising consumer loans. In fact, its use predated the introduction of Basel II, which merely standardised using 90 DPD as one of two possible default criteria. The other criterion still leaves some discretion to the lender, subject to regulatory approval. This is evident from paragraph 452 of \citet[pp.~100]{basel2006}, which defines `default' as the point at which the bank considers, \textit{in its opinion}, that the obligor is unlikely to repay its obligations in full, without the necessary intervention of the bank, e.g., liquidating any collateral. Basel II also lists six reasonable indicators of \textit{"unlikeliness to pay"} (or low repayment probability) in paragraph 453, e.g., having partially sold off the debt at a loss; or having restructured the agreement in such a way that the overall financial obligation is reduced.

After its introduction, many regulators adopted Basel II almost \textit{verbatim}, e.g., Regulation 67 of the Banks Act of \citet[pp.~1201--1202]{banksact1990}. Additionally, this Act indirectly describes `default' when defining \textit{non-performing debt} as having reached a point when it is \textit{"no longer prudent to credit interest receivable to the income statement"} of a bank. This definition is similar to the probabilistic idea of reaching some "point of no return", i.e., low repayment probability. In contrast, the recent guidelines (D403 of 2017) published by \citet{bcbs2017d403} intends to harmonise default definitions globally by proposing that 90 DPD be used as a universal definition of a non-performing loan. This was perhaps prompted by the European Banking Authority (EBA) having recently issued stricter guidelines on interpreting and applying Basel's default definition within the EU jurisdiction. These stricter versions of Basel's six default indicators are found in \citet{eba2016default}, which amends Article 178 of EU Regulation No 575/2013, also known as the Capital Requirements Regulation (CRR).

Capital aside, by forfeiting a portion of monthly income to cover \textit{expected} credit losses over the long run, the lender effectively caters for "business-as-usual" write-offs in principle, as explained in \citet[pp.~38--44]{VanGestel2009book} and \citet[pp.~167--169]{finlay2010book}. However, the recently introduced \citet{ifrs9_2014} accounting standard outlines additional principles specific to modelling these expected losses. IFRS 9 made this provisioning process generally more comprehensive and challenging when compared to its predecessor (IAS 39), as discussed in \citet{skoglund2017} and \citet{cohen2017new}. Regardless, IFRS 9 does not rigidly prescribe a single default definition. Instead, paragraph B.5.5.37 simply requires that a particular definition be consistently used throughout the relevant risk models and overall risk management. This requirement was recently enforced in \citet[\S~2.6]{eba2016default} for capital modelling in EU markets. The guidelines generally seek to align reaching Stage 3 credit impairment under IFRS 9 as a broader sign of `default', in harmonising some modelling elements between IFRS 9 and Basel. However, the guidelines concede that some differences in default definitions may be unavoidable, e.g., different legal entities within the same banking group, or different jurisdictions entirely. 

IFRS 9 allows for a rebuttable presumption of 90 DPD as a default definition, provided that an alternative definition be "\textit{reasonably}" demonstrated. In fact, Basel II previously allowed using default criteria of up to 180 DPD for retail and public sector exposures in paragraph 452, regulator-willing. This concession, as enshrined in Article 178(1)(b) of the CRR for EU markets, recognises that the quantum of this particular default criterion may differ amongst banks, presumably in line with each bank's risk tolerances amidst competition and market conditions. However, the EBA recently announced withdrawing this concession, primarily since only a small number of UK banks (and one French institution) are currently using it. The 2017/17 opinion piece of the \citet{eba2017dpd180}, annexed in \citet{eba2017dpd180annex}, argued that this withdrawal will harmonise reporting and remove \textit{"unwarranted"} variability in risk-weighted assets (RWA) across EU banks. Its analysis was mainly based on implied changes in RWA when using 90 DPD instead of 180 DPD as the default criterion. Using highly aggregated data from affected institutions, the EBA found that capital will likely increase for two thirds of these institutions. However, the analysis assumed all other factors will remain equal and largely ignored any fundamental opportunity costs and benefits at play when varying the `default' point, outside of RWA-centric capital modelling. Regardless, the UK regulator enforced this opinion recently in \citet{pra2019default} starting 31 December 2020.

The recent regulatory drive for standardising default definitions across jurisdictions is sensible from a compliance and comparability perspective. However, in doing so, regulators remove the rather probabilistic element within the original premise of a default definition, i.e., low repayment probability. They do this by decreeing certain criteria, most notably the 90 DPD threshold, to be risk `absolutes' beyond reproach. Instead of finding this threshold statistically using decision theory for each portfolio, a standardised default threshold devolves into little more than a static hurdle. In contrast hereof, consider an operational perspective wherein reaching `default', which is fundamentally defined as the probabilistic "point of no return", is impetus for the lender to act, e.g., initiating stringent debt recovery proceedings and abandoning the credit relationship. Surely there must be varying consequences associated with the lender's action in this regard and, more importantly, its exact timing. If this is remotely true, then modelling any aspect of the `default' event using potentially stale definitions thereof will likely be sub-optimal. As paraphrased from \citet{hand2001m}, it is reasonable to question the pursuit of modelling excellence when the constructed outcome variable itself, as constrained by the default definition, is inherently quite arbitrary.

\subsection{Delinquency and default definitions: a servant of many masters}
\label{sec:2_default_servant_many_masters}

Default definitions serve a few other masters beyond that of modelling regulatory capital or provisions for unexpected and expected losses respectively. They are also used when building application credit scorecards, which are automated decision-making models rendering consistent approve/decline credit decisions based on estimated default risk, as discussed in \citet{hand1997}, \citet{hand2001m}, \citet[chap~1,~8]{thomas2002credit},
\citet{siddiqi2005credit}, \citet[chap~1]{thomas2009book},
\citet[chap~4]{VanGestel2009book}, \citet{thomas2010consumer}, and \citet{louzada2016classification}. Banks generally prefer a decent risk-ranking ability over attaining maximal predictive accuracy in scorecards, which is perhaps characteristically pragmatic. To this end, finding a suitable definition is a largely subjective and iterative exercise, wherein a few qualitative factors are balanced. When setting $d$ itself, the type of security/collateral is important given the different workout experiences in debt recovery, e.g., mortgages versus personal loans. Moreover, $d$ cannot sensibly exceed the maximum contractual loan term, which is especially relevant for very short term products such as payday loans. However, the bank's risk appetite is perhaps the most important factor: the higher the appetite for arrears (compensated by higher interest income), the greater $d$ will be, and \textit{vice versa}, as an intuitive principle.

Lenders conduct a form of statistical analysis, called a \textit{roll rate analysis}, in providing quantitative assurance on a chosen $d$, as outlined in \citet[pp.~41--42]{siddiqi2005credit}. This is a cross-tabulation of observed transition rates across a length of time amongst increasingly severe arrears categories (30 days, 60 days, etc.), including the newly-chosen "default state" as imposed via $d$. The principle is based on back-solving for stability in that accounts identified as `lost' should stay lost at the end of the outcome period. Having found this point, the resulting default definition using $d$ should ideally remain stable over time, with only a minimum of accounts recovering from default. However, this approach ignores the competing financial/opportunity costs that may be in play when varying $d$ itself, e.g., legal and administration costs, loss provision changes, salaries, and collection efforts. Instead, the direct loss implications associated with any $d$ may be a better criterion than stability. Another source of variation is the chosen epoch of time from which loan performance is sampled, which can shroud the `true' transition rates, and definitely influence the choice of $d$. However, perhaps the greatest uncertainty is that of the chosen outcome period, which can obscure certain idiosyncratic risk characteristics of a portfolio -- and therefore the `best' choice of $d$ -- simply due to choosing an inappropriate outcome period.

Although literature from \citet[pp.~99]{thomas2002credit} and \citet[pp.~101--102]{VanGestel2009book} suggests a typical range of 6--24 months, the effects of various outcome periods were more rigorously explored in \citet{kennedy2013window} and \citet{mushava2018experimental}. The authors predicted default risk across a few differing time spans using Irish and South African data respectively. Too short a window of time may be insufficient to capture the various transition rates due to a lack of maturity and/or seasonal effects. Conversely, too long a window may no longer represent either the current market conditions or the portfolio's current risk composition, and was shown to yield decreasingly accurate models. Too long a window may also ignore rapid movements amongst delinquency states, e.g., oscillating instances between defaulting and curing, as argued in \citet{kelly2016good}, which may especially affect monthly loss provisioning. Clearly, the outcome period and the sample window are significant factors that complicate choosing $d$. As an example, it would be difficult to decide if a particularly low curing rate given $d$ is artificially due to an overly short outcome period, testament of `true' risk, or shifting market conditions -- without conducting additional analysis. For these reasons, a roll rate-based approach is deemed unfit for finding $d$ dynamically in this study.

A basic Markov chain can be considered as a more advanced form of a roll rate analysis. This was first explored in \citet{cyert1962} for loss provisioning, and later refined in \citet{corcoran1978use} and \citet{vanKuelen1981note}, wherein a time-homogeneous discrete-space Markov chain modelled the transitions amongst arrears categories. The bankruptcy process of firms was modelled in \citet{jarrow1997markov} as a Markov chain across various credit rating states and an absorbing default state. Specifically, the time distribution of the chain first entering default was explicitly modelled, from which a PD was estimated over a certain term structure. A non-stationary forecasting model with a Markovian structure was developed in \citet{smith1995forecasting} using US mortgage loans. This was extended in \citet{grimshaw2011markov} using US subprime mortgages, wherein new empirical Bayes estimators for non-stationary and heterogeneous transition matrices were proposed. This heterogeneity recognises that certain segments within a loan portfolio may have fundamentally different transition rates, from which we draw some inspiration in the present study. A four-state non-homogeneous Markov chain was  built as part of a larger suite of intensity models in \citet{leow2014intensity} for credit card delinquencies. Clearly, Markov models are quite prolific in modelling credit risk, especially evident from the rich review given thereof in \citet{hao2010review}.

The mere possibility of an account recovering from `default' (as fixed via $d$) inherently introduces uncertainty in any chosen $d$ as the supposed "point of no return". Anecdotal experience corroborates this in that multi-period `episodes' of delinquency are more widespread in practice than one would otherwise believe. In fact, the work of \citet{thomas2016} demonstrated this oscillating regime-switching effect between payment and non-payment, having modelled the collections process of defaulted UK loans using both a four-state homogeneous Markov chain and a time-sensitive hazard rate model. These models may be indirectly used to test write-off policies by incorporating them as boundary conditions for loan duration, though the authors admitted this is not really an appropriate solution. Instead, \citet{mitchner1957operations} investigated the optimal pursuit duration of loan recovery based on maximising the net profit of a collections department. Using US personal loans, they found that loan recovery should cease whenever the one-period expected repayment equals the cost of pursuing recovery itself, which is quite intuitive. However, the authors admitted that their results depend critically on the assumption that a defaulted borrower is forever absorbed into a paying regime, once entered, which contrasts \citet{thomas2016}.

In optimising the collections process, \citet{de2010optimizing} devised a dynamic programming model wherein both the ideal type of recovery action and its pursuit duration are found every month for the "average" debtor. Using unsecured European loans, their objective was to maximise the net recovery rate by pursuing a particular action for a set period, including telephonic calls, various letters, house visits, threats, legal redress, and write-off. However, the state space formulation excludes cash flows themselves from previous periods or indeed those from the future. Their work was later extended in \citet{so2019debtor} wherein a Bayesian approach was followed to obtain similarly optimised outputs, though on the individual debtor-level. Within the same context, the work of \citet{liu2019markov} devised a Markov-based decision process wherein the optimal collection action is theoretically found across both delinquency state progressions and time, using designed data. A schedule of optimal collection actions is then calculated accordingly, which was shown to supersede a static policy based on maximising the expected net present value. However, the authors made some strong simplifying assumptions when designing both their `data' and elements within their method, which may have implications for their end-results when trying to use their method on real data. In addition, they impose explicit write-off criteria exogenously within the state space of their Markov chain, instead of structuring it as a candidate collection action.

Another extension is that of \citet{duman2017novel} wherein outputs similar to that of \citet{de2010optimizing} were sought using Turkish data, though on a finer time scale using days. The authors posited that each recovery action may have unintended consequences on subsequent customer loyalty, which they incorporated into their optimisation along with collection capacity constraints. \citet{chehrazi2019dynamic} examined the same collection problem and formulated repayments as a complex self-exciting point process in continuous time. As a stochastic control problem, the authors allowed both the size and the timings of receipts to influence each other, which in turn are perturbed by pursuing a particular recovery action. Another relevant contribution is that of \citet{matuszyk2010modelling} wherein a decision tree-based approach was developed to help inform the optimal collection strategy. Candidate choices included collecting in-house, outsourcing collection efforts externally, or selling the debt. Their work formed part of subsequently modelling the LGD using a two-stage approach and unsecured personal loans. Lastly, \citet{han2013effects} extended this by showing the positive effects of including collection action history within LGD models, using Korean data.

The present study is closest in form to that of \citet{de2010optimizing} and \citet{liu2019markov}, though we follow a different and more general approach based on delinquency measures that leverages the entire portfolio instead of only defaulted loans. Moreover, we focus more fundamentally on if and when to forsake a loan based on delinquency progression over time, instead of attempting to compile a menu of collection actions. The idea of varying thresholds $d$ within a scorecard's default definition was actually first investigated in \citet{harris2013default} and \citet{harris2013quantitative}, in which the effects on classifier accuracy (using support vector machines) were studied. However, those findings -- while proving that $d$ significantly influences model accuracy -- ignore the direct loss implications associated with a particular $d$. Amongst other things, our work attempts to bridge some gaps between the collection and credit risk modelling literatures. This is demonstrated by loss-optimising the recovery decision's timing as the "point of no return" in line with the original notion underlying a default definition.

\section{Two forecasting techniques for loss-optimising the recovery decision based on measured delinquency}
\label{sec:3}

Three delinquency measures were formulated in our previous work in \citet{botha2020paper1}: the commonly-used \textit{Contractual Delinquency} $g_1$ as the weighted number of monthly payments in arrears, which is also the main measure used in this study; the \textit{Macaulay Duration}-based $g_2$ as an index of the weighted average time to recover the capital portion; and the \textit{Degree of Delinquency} $g_3$ as a modified variant of $g_2$ that incorporates the sizes of disrupted cash flows in assessing delinquency. Constructing $g_1$ relies on calculating the repayment ratio between a receipt and an instalment at a particular period $t$, denoted as $h_t$. Delinquency only increases if $h_t$ is below a given level, which is set to 90\% as in our previous work. Finally, the LROD-procedure requires three preparatory steps:\begin{enumerate}
	\item Delinquency must be measured for every account across its history using a measure $g\in \{g_1,g_2,g_3\}$;
	\item Select appropriate thresholds $d \in \mathcal{D}_g$ on the domain of a particular $g$ for optimisation;
    \item A portfolio loss model $L_g$ must be applied at every chosen  threshold $d \in \mathcal{D}_g$ of $g$.
\end{enumerate} 

Given a threshold $d\geq 0$ on the domain of $g$, denoted henceforth as $(g,d)$, the main optimisation problem is effectively divided into smaller $(g,d)$-based sub-problems by storing each loss estimate $L_g(d)$ into a central collection. A threshold $d'\in\mathcal{D}_g$ is then sought such that $L_g(d')\leq L_g(d)$ for all chosen $d\in\mathcal{D}_g$. Finally, a global minimum loss $m^{(g)}$ may exist at a threshold $d^{(g)}$ for a measure $g$, which is formally expressed as \begin{align} 
        m^{(g)} &= \min_{d\in\mathcal{D}_g}{ L_g(d)} \quad \text{and} \label{eq:loss_proc_a} \\
        d^{(g)} &= \arg_d\,\min_{d\in\mathcal{D}_g}{ L_g(d)} \, . \label{eq:loss_proc_b}
\end{align} If competing measures are used, then the optimal measure $g^{\ast}$ is the function $g$ that yielded the lowest losses amongst $\left[m^{(g_1)}, m^{(g_2)}, m^{(g_3)}\right]$, thereby making delinquency measures comparable to one another in the aggregate. This concludes the LROD-procedure itself.

Similar to our previous work, the portfolio of $N$ loans is segmented into two subsets of accounts, denoted as $\mathcal{S}_D$ and $\mathcal{S}_P$, for a particular $(g,d)$-configuration. Each account $i \in \mathcal{S}_D$ is considered as $(g,d)$-defaulting if and only if $g(i,t)\geq d$ at any particular time $t=0,\dots,t_c$ during its history, where $t_c$ is that loan's full contractual term. If this condition does not hold, then the account is logically considered as $(g,d)$-performing and contained within the other set $\mathcal{S}_P$. Every $(g,d)$-defaulting account has an earliest moment $t_i^{(g,d)}$ of entering the `default' state, depending on the particular $(g,d)$-configuration. Note that `default' is a contrived and more fundamental state in this study, which is best interpreted as one possible end point of debt recovery, given $(g,d)$. The same model $L_g$ is used, i.e., the segmented sum of all discounted account-level losses $l(i,t_i)$ is calculated across all accounts $i=1,\dots,N$, as assessed from a particular time point $t_i$ of the loan. This sum is split between $(g,d)$-defaulting and $(g,d)$-performing loans that are either assessed back from $t_i^{(g,d)}$ or from the contractual loan term $t_{c_i}$, formally defined as \begin{equation} \label{eq:total_loss}
	L_g = \sum_{i \in \ \mathcal{S}_D} {l\left(i,t_i^{(g,d)} \right)} \ + \ \sum_{i \in \ \mathcal{S}_P} {l\left(i,t_{c_i}\right)} \ .
\end{equation} The discounted loss itself $l(i,t_i)$ is the weighted sum of the expected outstanding balance and the arrears amount, both observed at a given time $t_i$ and weighted respectively by the loss rates $r_E=40\%$ and $r_A=70\%$. These quantities are discounted using a risk-free rate of 7\%, which is realistic for the South African market. Using two different loss rates recognises that the recovery success may differ between these two components. These rates are chosen arbitrarily such that losses on arrears ought to be penalised more than losses on expected balances in principle. They are intended as placeholders for the output from more sophisticated loss models, which presumably includes all other costs.

However, differing from the previous study, the receipt vector $\boldsymbol{R}=\big[R_1,R_2,\dots,R_{t_0},R_{t_1},\dots,R_{t_c}\big]$ of a real loan is only observable from data up to time $t_0 \leq t_c$, owing to the right-censoring effect that is inherent to most credit portfolios in practice. The smaller the difference $t_c-t_0$, the less treatment such a loan would need in completing its history up to contractual maturity, and \textit{vice versa}. Not treating a loan portfolio in this way will likely lead to unusable and unsatisfying results when applying the LROD-procedure (see appendix). As such, two techniques are devised for treating an empirical loan portfolio by forecasting the receipts $R_{t_1},\dots,R_{t_c}$ up to the contractual term $t_c$ for each loan account, using its observed receipt history $R_1,\dots,R_{t_0}$. These include a simple probabilistic technique called \textit{random defaults} as well as a more sophisticated eight-state Markov chain-based technique called \textit{Markovian defaults}.

\subsection{Random defaults with empirical truncation}
\label{sec:3-random_defaults}

Let $u_t\in [0,1]$ be a randomly generated number at every loan period $t=t_1,\dots,t_c$ to be forecast and let $b$ be an estimable probability of payment, i.e., $P(R_t=I_c) = b$ with $I_c$ denoting the calculated level instalment. Note that $I_c$ is the instalment that amortises the outstanding balance at time $t_0$ to zero at time $t_c$, using the latest observed client interest rate. The receipt is then iteratively forecast at each future $t$ as \begin{equation} \label{eq:random_defaults}
	R_t = \begin{cases}
    		I_c & \text{if} \ u_t < b \\
            0 & \text{otherwise}
    	  \end{cases} \ .
\end{equation} 

A randomised truncation effect is introduced (similar to our previous work) via a structural break in the forecast receipt vector at a certain point (if at all) and replacing elements thereafter with zeros. Tempering the forecast receipts in this way mimics the fact that some loan accounts will simply never resume payment in reality. This is similar to \citet{thomas2016} wherein the parameters controlling the payment and non-payment sequences were fixed after reaching some point in the process. More formally, consider all periods $j=t_0\,,\dots,t_c$ within the now-forecast receipt vector $\boldsymbol{R}$ of a particular account, with the measure $g_1$ applied accordingly across all periods. Let $k\geq 0$ be a truncation parameter above which the receipts are truncated. The starting period of this truncation, denoted as $t_k \geq 0$, may then exist if the account has experienced sufficient delinquency $g_1(j)\geq k$ at some $j$, i.e., $t_k= \min\big(\, j : g_1(j) \geq k\big)$. Conversely, if delinquency has not breached $k$, then this time point $t_k$ does not exist. A process called $(k,g_1)$-truncation then changes $\boldsymbol{R}$ to $\boldsymbol{R}'$ by \begin{equation} \label{eq:truncation}
	\boldsymbol{R}' = \begin{cases}
               \big[R_{t_1},\dots,R_{t_k},0 \,,\dots,0\big]  & \text{if} \ t_k \ \text{exists} \\
               \boldsymbol{R} & \text{otherwise}
            \end{cases} \ .
\end{equation} In estimating this truncation parameter $k$, consider that the maximum delinquency across time can be obtained for each account in a loan portfolio, using $g_1$ for simplicity's sake. In turn, the histogram of these maxima is plotted, followed by fitting statistical distributions to these maxima. One can then draw a random sample $\hat{k}_i$ from an appropriately fitted distribution for each account and finally $(\hat{k}_i,g_1)$-truncate the initially forecast receipt vector. This introduces a bit of realistic variance to the overall truncation effect.

Lastly, consider an indicator function $\mathcal{I}_{t}^{\,(i)}$ that signals payment using the receipt $R_t^{\,(i)}$ and instalment $I_t^{\,(i)}$ of the $i^{\text{th}}$ account at historical periods $t=1,\dots,t_0(i)$ where $t_0(i)$ denotes the most recently observed loan period for this account. This is then formally defined as \begin{equation} \label{eq:ind_pn}
    \mathcal{I}_{t}^{\,(i)} = \begin{cases}
                 1 & \text{if} \ R_t^{\,(i)}\geq I_t^{\,(i)} \\
                 0 & \text{otherwise}
                  \end{cases} \quad t=1,\dots,t_0(i) \ .
\end{equation} Accordingly, the probability of payment $b$ is estimable by $\hat{b}$, defined as \begin{equation} \label{eq:b_estimator}
    \hat{b} = \frac{1}{N}\sum_i{\frac{1}{t_{0}(i)}\sum_t{\mathcal{I}_{t}^{\,(i)}}} \quad \forall \ i=1,\dots,N \ \text{accounts and} \ t=1,\dots,t_0(i) \ \text{periods} \ .
\end{equation}

\subsection{Markovian defaults}
\label{sec:3-markovian_defaults}

Let $X_t\in \{x_0,\dots,x_7\}$ be a random vector that can assume one of eight increasingly-severe delinquency states derived from $g_1$, across all historical periods $t=1,\dots,t_0$ of an account. The states $x_0\,,\dots,x_5$ correspond to $g_1(t)$ having the respective values $0,\dots,5$ at any $t$. State $x_6$ is semi-absorbing such that $g_1(t)\geq 6$ at any $t$ and the state $x_7$ denotes write-off (fully-absorbing). The sequence $X_1,\dots,X_{t_0}$ then forms a discrete-time first-order Markov chain from which receipts can be forecast, based on the predicted states $X_{t_1},\dots,X_{t_c}$ at future periods $t=t_1,\dots,t_c$. Note that $g_1$ can only ever increase in value by one delinquency level, while it can decrease by several levels depending on the magnitude of the overpayment $R_t>I_t$. 

To generate receipts at these future periods, temporarily ignore write-off ($x_7$) and consider the one-period delinquency difference $\delta_t$, defined as $\delta_t = g_1(t) - g_1(t-1)$. A positive difference $\delta_t>0$ implies $R_t < h_t I_c$ since delinquency has increased and $R_t$ is therefore simply zeroed. Secondly, $\delta_t=0$ implies $R_t=I_c$ since the delinquency level remained unchanged. Finally, $\delta_t<0$ implies $R_t \geq 2I_c$ since $\delta-1$ extra payments are needed to decrease the delinquency level beyond the instalment normally due at the time. When $X_t=x_6$, the account remains semi-absorbed as long as $g_1(t)\geq 6$, which implies either increasing or constant delinquency. For the sake of prudence, the former case is assumed (i.e., $\delta_t>0$) and $R_t$ is zeroed accordingly. These ideas (barring $x_7$) are combined into forecasting the receipt as \begin{equation} \label{eq:markov_defaults}
    R_t = \begin{cases}
            -I_c(\delta_t-1) & \text{if} \ \delta_t<0  \\ 
            I_c & \text{if} \ \delta_t=0  \\
            0                & \text{if} \ \delta_t > 0
          \end{cases} \ .
\end{equation} Note that truncation is effectively incorporated whenever an account transitions to the write-off state $x_7$ at a supposed time point $t_w$ that only exists when $X_{t_w}=x_7$ with $t_1 \leq t_w \leq t_c$. This implies zeroed receipts from that point forward, i.e., $R_t=0$ for $t=t_w,\dots,t_c$ if $t_w$ exists.

Regarding the transition matrix of this Markov chain, note that the receipt history of each loan account signifies a repeated observation of the underlying chain, as discussed in \citet{anderson1957statistical}. Assuming stationarity, the maximum likelihood estimates (MLEs) for the transition probabilities $p_{ij}$ from state $i$ to state $j$ are then $\hat{p_{ij}}=n_{ij}/n_i^\ast$ where $n_{ij}$ is the number of observed transitions across all time periods from state $i$ to $j$ and $n_i^\ast$ is the observed number of total transitions starting in state $i$. It is not necessary to estimate initial state probabilities since the starting delinquency state is simply observed from the last available time point $t_0$ of an account.
\section{Calibrating the forecasting techniques to mortgage data}
\label{sec:4}

Short term matured loans would be ideal for this study, though only mortgage data was available; specifically, a rich portfolio of ordinary home loans granted to the lower-income segment of the South African market. This longitudinal dataset has monthly loan performance observations over time $t=1,\dots,t_0(i)$ for account $i=1,\dots,N$ with $N=61,648$ single-advance 20-year mortgage accounts. Mortgages originated from April 2004 (and beyond) are extracted and observed up to December 2017, thereby yielding $3,271,534$ raw monthly observations of loan performance. The data itself includes actual cash flows (receipts), expected instalments (including credit life insurance add-ons and fees, or special arrangements), variable interest rates, original loan principals, month-end balances, write-off amounts, asset sale proceeds, and early settlement indicators.

\begin{figure}[ht!]
\centering\includegraphics[width=0.8\linewidth,height=0.47\textheight]{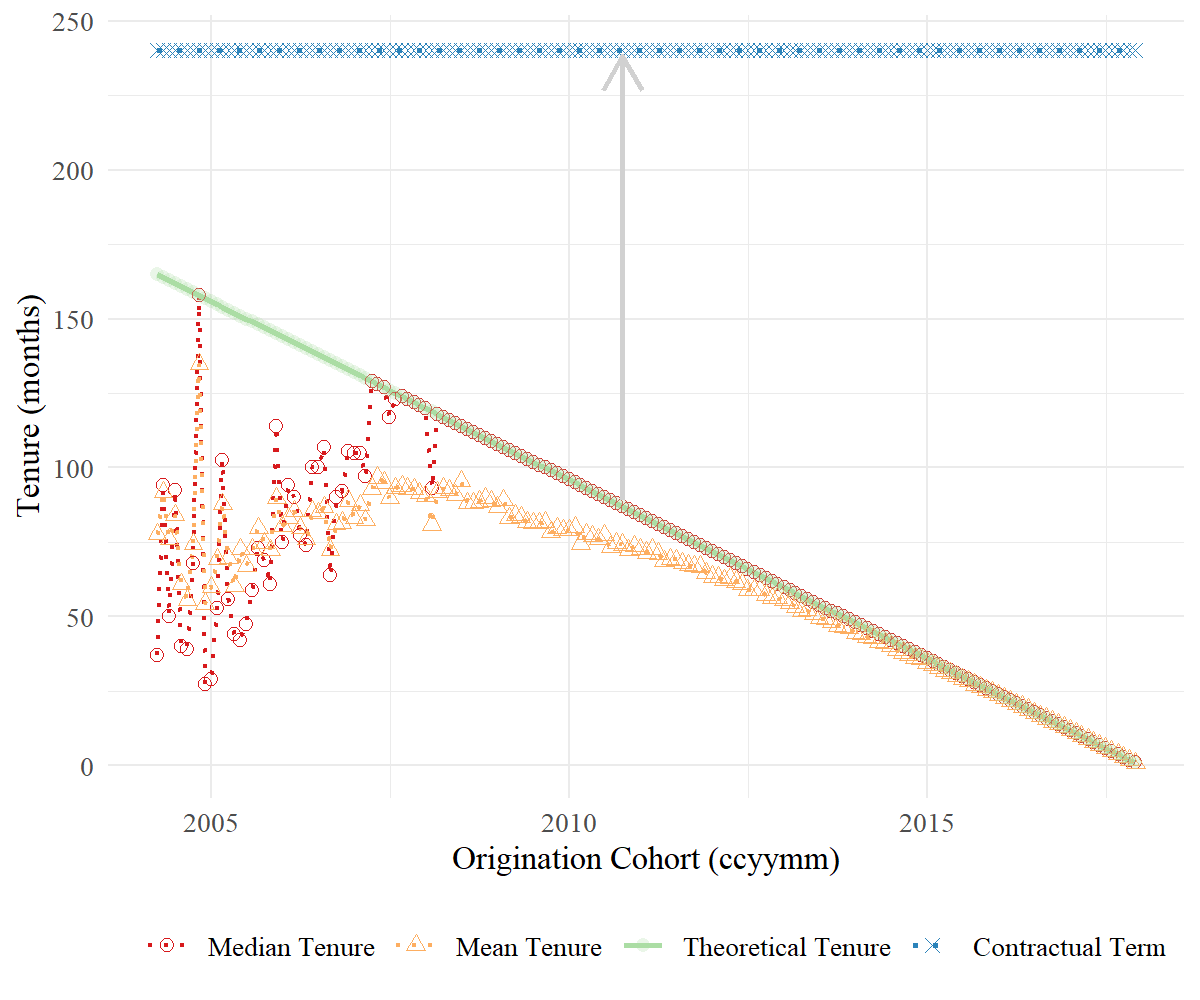}
\caption{The difference between the theoretically observable loan tenure (as measured at December 2017 in retrospect) and the remaining contractual term across monthly loan cohorts. The mean and median loan ages per monthly cohort are overlaid.}\label{fig:EmpiricalLoanTenures}
\end{figure}

Similar to \autoref{fig:MainObjective}, the difference between the maximum theoretical loan tenure and the remainder of the contractual term is shown in  \autoref{fig:EmpiricalLoanTenures} at every historical monthly loan cohort, with aggregates overlaid. The fact that these aggregates are below the theoretical maximum for most cohorts demonstrates some additional right-censoring. In particular, mortgage loans can exit the portfolio pre-maturely either via write-off or via early settlement, e.g., private sales, bond cancellations, or transfers. The volatility in both aggregates at earlier times attests to low sample sizes, which is unsurprising for a fledgling loan portfolio at the time. This volatility gradually subsides until both aggregates approach the theoretical maximum. This is sensible as more recently originated mortgages have less time available to develop write-off/settlement outcomes than their older counterparts.

\begin{figure}[ht!]
\centering\includegraphics[width=0.47\linewidth,height=0.32\textheight]{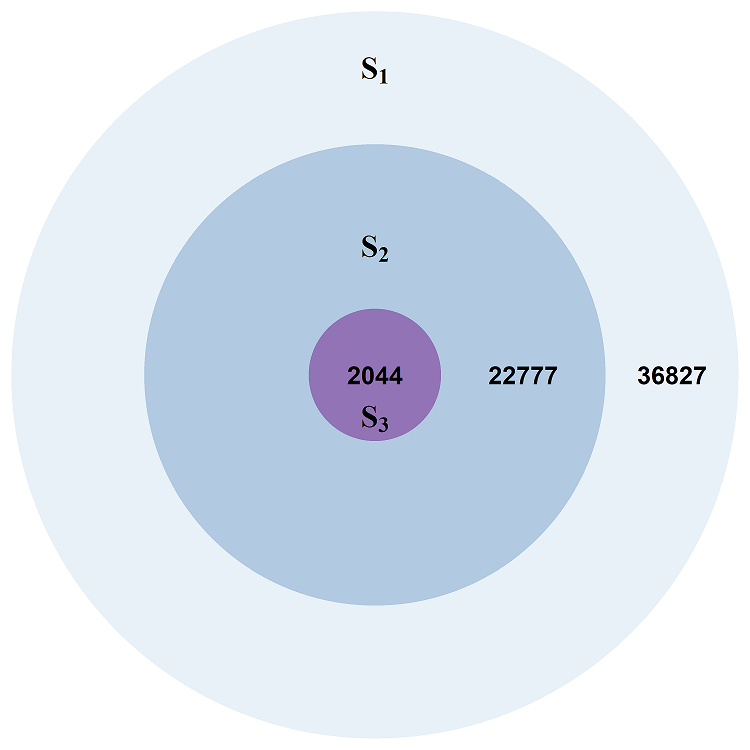}
\caption{A Venn diagram showing the relative sizes and overlaps amongst the three main samples of mortgage accounts: $S_1$ (full sample), $S_2$ (delinquents), and $S_3$ (write-offs). These samples are used both in training the forecasting techniques and during the subsequent loss optimisation.}\label{fig:VennSamples}
\end{figure}

In estimating the various parameters of the forecasting techniques, the data is partitioned to form three specific samples: $S_1$ as the full dataset, $S_2$ as the delinquents-only sample (all accounts that had at least one payment in arrears historically, or were eventually written-off), and $S_3$ as the write-offs sample. These samples and the relationships amongst them are illustrated with a Venn diagram shown in \autoref{fig:VennSamples}. Some accounts will simply never experience any delinquency and their exclusion in $S_2$ and $S_3$ removes an optimism bias during model training. There is little practical benefit to finding the best time for loan recovery on a near risk-less portfolio. Furthermore, recovery optimisation is only sensible for loans likely to become delinquent in the first place, which is predicated upon forecasting them as such. Likewise, it would be pointless to forecast cash flows of closed accounts, though their repayment histories are retained for optimisation purposes. Lastly, this particular partitioning scheme is an experimental proxy for risk compositions that differ across both product and risk appetites in reality. As an example, mortgages typically have a much lower default rate than unsecured personal loans, which is catered for in our setup.

\subsection{Calibrating the random defaults technique}
\label{sec:calib_random_defaults}

\begin{figure}[ht!]
\centering\includegraphics[width=0.8\linewidth,height=0.47\textheight]{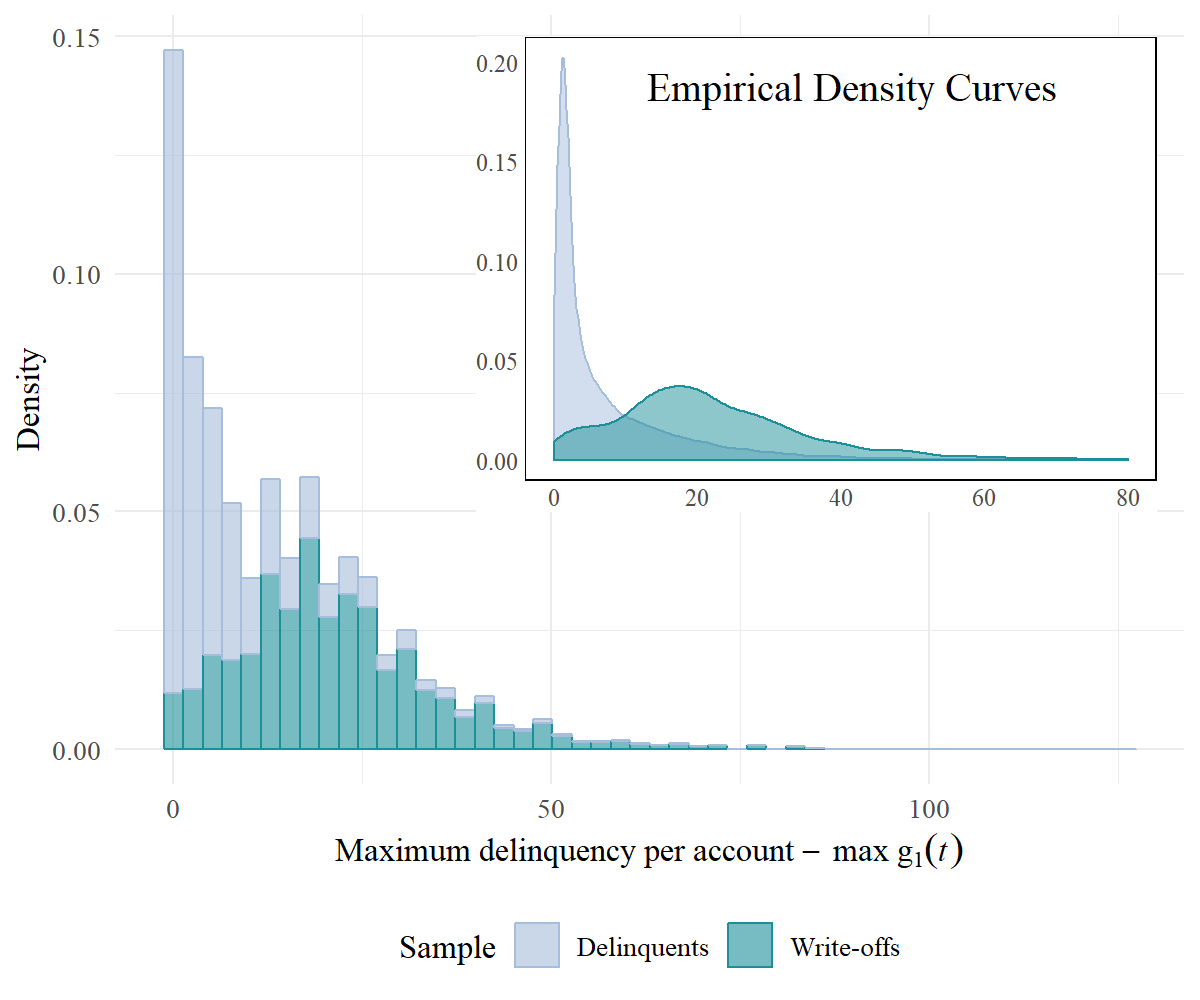}
\caption{A histogram and empirical density curve of the maximum delinquency level observed per account, drawn for the samples $S_2$ (delinquents) and $S_3$ (write-offs). A theoretical distribution is then fit on each sample (see appendix), from which the truncation parameter $k$ is drawn randomly for each loan account prior forecasting. }\label{fig:Dist_Max_CD}
\end{figure}

The aforementioned probability of payment $b$ used in this technique is estimated from samples $\{S_1, S_2, S_3\}$ respectively as $\hat{b}_1=87\%$, $\hat{b}_2=81\%$, and $\hat{b}_3=45\%$. The descending values are plausible given that each successive sample contains a greater proportion of delinquency by design. Note that the random truncation of forecasts is only sensibly performed for delinquent cases, which implies $k>0$. Therefore, ignoring $S_1$, the distribution of the maximum delinquency level per account, i.e., $\text{max}\,g_1(t)$ across all historically observed periods $t=1,\dots,t_0$, is given in \autoref{fig:Dist_Max_CD} for both samples $S_2$ and $S_3$. A few statistical distributions were tested against the data (see appendix), though the exponential and two-parameter Weibull distributions are chosen for $S_2$ and $S_3$ respectively, denoted as $\text{Exp}(\lambda)$ and $\text{Weibull}(\lambda,\phi)$. The MLEs of these parameters are $\lambda=0.1378555$ for the exponential distribution, scale $\lambda=24.449566$ and shape $\phi=1.688026$ for the Weibull distribution. The truncation parameter then follows either one of these distributions, i.e., $k\sim\text{Exp}(\lambda)$ for both $S_1$ and $S_2$, as well as $k\sim\text{Weibull}(\lambda,\phi)$ for $S_3$, as part of a comparative study. Note that the exponentially-distributed $k$ has a stronger truncation effect since it generally yields lower values of $k$ than those yielded by its Weibull-distributed counterpart. This is also evidenced by the sample mean of $k$ estimated from $S_2$ being $7.25$ versus that from $S_3$ being $21.58$.

\subsection{Calibrating the Markovian defaults technique}
\label{sec:calib_Markovian_defaults}

The MLEs for the transition matrix used in this technique are estimated only from the samples $S_2$ and $S_3$, shown respectively in \crefrange{tab:markov1}{tab:markov2}. Note that the estimates using $S_1$ differ from those yielded by using $S_2$ only in the first row, which is sensible since $S_1$ contains the same delinquent accounts (and therefore the same transitions) as $S_2$ by design. The estimates are realistic in that an account in any given delinquency state (barring write-off) can increase its delinquency only by one level within a monthly period. Additionally, these estimates reflect the fact that an account can significantly overpay and thereby recover either partially or entirely from distress. The probability of staying within a starting state is greatest, though it decreases gradually as the delinquency level increases, at least for states $x_0,\dots,x_5$. Simultaneously, the probability of becoming even more delinquent increases as the starting delinquency level increases, which agrees with anecdotal experience. This is corroborated by the increasing probability of write-off, effectively representing an increasing probability of truncation.

\begin{table}[ht!]
\centering
\begin{tabular}{@{}llllllllll@{}}
\multicolumn{1}{c}{} & \multicolumn{9}{c}{Ending state} \\
 &  & \multicolumn{1}{c}{$x_0$} & \multicolumn{1}{c}{$x_1$} & \multicolumn{1}{c}{$x_2$} & \multicolumn{1}{c}{$x_3$} & \multicolumn{1}{c}{$x_4$} & \multicolumn{1}{c}{$x_5$} & \multicolumn{1}{c}{$x_6$} & \multicolumn{1}{c}{$x_7$}  \\ \cline{3-10}
\multirow{9}{*}{\rot{Starting state}}  & \multicolumn{1}{r|}{$x_0$} & 0.9477 & 0.0521 & 0.0000 & 0.0000 & 0.0000 & 0.0000 & 0.0000 & \multicolumn{1}{l|}{0.0002}\\
 & \multicolumn{1}{r|}{$x_1$} & 0.0942 & 0.8074 & 0.0980 & 0.0000 & 0.0000 & 0.0000 & 0.0000 & \multicolumn{1}{l|}{0.0004}\\
 & \multicolumn{1}{r|}{$x_2$} & 0.0138 & 0.0502 & 0.7735 & 0.1621 & 0.0000 & 0.0000 & 0.0000 & \multicolumn{1}{l|}{0.0004}\\
 & \multicolumn{1}{r|}{$x_3$} & 0.0064 & 0.0084 & 0.0481 & 0.7372 & 0.1993 & 0.0000 & 0.0000 & \multicolumn{1}{l|}{0.0006}\\
 & \multicolumn{1}{r|}{$x_4$} & 0.0064 & 0.0030 & 0.0082 & 0.0488 & 0.6957 & 0.2371 & 0.0000 & \multicolumn{1}{l|}{0.0007}\\
 & \multicolumn{1}{r|}{$x_5$} & 0.0051 & 0.0020 & 0.0029 & 0.0081 & 0.0469 & 0.6846 & 0.2496 & \multicolumn{1}{l|}{0.0009} \\
 & \multicolumn{1}{r|}{$x_6$} & 0.0044 & 0.0006 & 0.0007 & 0.0009 & 0.0021 & 0.0095 & 0.9756 & \multicolumn{1}{l|}{0.0061} \\
 & \multicolumn{1}{r|}{$x_7$} & 0.0000 & 0.0000 & 0.0000 & 0.0000 & 0.0000 & 0.0000 & 0.0000 &  \multicolumn{1}{l|}{1.0000} \\ \cline{3-10} 
\end{tabular}
\caption{Maximum likelihood estimates for the transition matrix of the multi-state Markov chain, estimated from the delinquents sample $S_2$. States $x_0,\dots,x_5$ correspond to $g_1$ having the respective values $0,\dots,5$ (weighted payments in arrears). States $x_6$ (semi-absorbing) and $x_7$ (absorbing) indicate $g_1\geq 6$ and write-off respectively.} \label{tab:markov1}
\end{table}

\begin{table}[ht!]
\centering
\begin{tabular}{@{}llllllllll@{}}
\multicolumn{1}{c}{} & \multicolumn{9}{c}{Ending state} \\
 &  & \multicolumn{1}{c}{$x_0$} & \multicolumn{1}{c}{$x_1$} & \multicolumn{1}{c}{$x_2$} & \multicolumn{1}{c}{$x_3$} & \multicolumn{1}{c}{$x_4$} & \multicolumn{1}{c}{$x_5$} & \multicolumn{1}{c}{$x_6$} & \multicolumn{1}{c}{$x_7$}  \\ \cline{3-10}
\multirow{9}{*}{\rot{Starting state}} & \multicolumn{1}{r|}{$x_0$} &  0.8820 & 0.1126 & 0.0000 & 0.0000 & 0.0000 & 0.0000 & 0.0000 & \multicolumn{1}{l|}{0.0054} \\
 & \multicolumn{1}{r|}{$x_1$}  & 0.0962 & 0.5387 & 0.3534 & 0.0000 & 0.0000 & 0.0000 & 0.0000 & \multicolumn{1}{l|}{0.0117} \\
 & \multicolumn{1}{r|}{$x_2$} & 0.0254 & 0.0453 & 0.4607 & 0.4600 & 0.0000 & 0.0000 & 0.0000 & \multicolumn{1}{l|}{0.0086} \\
 & \multicolumn{1}{r|}{$x_3$} & 0.0136 & 0.0103 & 0.0430 & 0.3824 & 0.5393 & 0.0000 & 0.0000 & \multicolumn{1}{l|}{0.0114} \\
 & \multicolumn{1}{r|}{$x_4$} & 0.0117 & 0.0032 & 0.0093 & 0.0412 & 0.3187 & 0.6048 & 0.0000 & \multicolumn{1}{l|}{0.0112} \\
 & \multicolumn{1}{r|}{$x_5$} & 0.0079 & 0.0037 & 0.0037 & 0.0053 & 0.0293 & 0.3181 & 0.6194 & \multicolumn{1}{l|}{0.0127} \\
 & \multicolumn{1}{r|}{$x_6$} & 0.0076 & 0.0006 & 0.0005 & 0.0007 & 0.0012 & 0.0035 & 0.9474 & \multicolumn{1}{l|}{0.0385} \\
 & \multicolumn{1}{r|}{$x_7$} & 0.0000 & 0.0000 & 0.0000 & 0.0000 & 0.0000 & 0.0000 & 0.0000 &  \multicolumn{1}{l|}{1.0000}  \\  \cline{3-10} 
\end{tabular}
\caption{Maximum likelihood estimates for the transition matrix of the multi-state Markov chain, estimated from the write-offs sample $S_3$. States $x_0,\dots,x_5$ correspond to $g_1$ having the respective values $0,\dots,5$ (weighted payments in arrears). States $x_6$ (semi-absorbing) and $x_7$ (absorbing) indicate $g_1\geq 6$ and write-off respectively.} \label{tab:markov2}
\end{table}

\subsection{Assessing the quality of forecasts}
\label{sec:calib_accuracy}

Although forecasts are trained specifically on $\left\{S_1, S_2, S_3 \right\}$, the forecast quality itself is examined herein by following a more general $k$-fold cross-validation approach as additional assurance. However, our particular mortgage portfolio did not have a single completed 20-year loan, against which the receipt forecasts could be validated across \textit{all} periods. Nonetheless, available loan data up to $t_0(i)$ is still used within a $k=5$ setup, despite the censoring-related bias this likely introduces into measuring the forecast error. Moreover, our main objective is not to produce the most accurate or robust forecasting model on the account-level, although that is certainly a worthwhile endeavour. Instead, we focus on the more fundamental effect of using different forecasts when optimising the timing of loan recovery, which by implication agrees with using multiple forecasting techniques.

\begin{table}[ht!]
\centering
\begin{tabular}{@{}lll@{}}
\toprule
Metric & Random defaults ($T_a$) & Markovian defaults ($T_b$) \\\midrule
Portfolio Arrears Rate (PAR)                          & 6.715\%      &   0.695\%                                \\
Mean parameter \%-difference                          & 0.00012\%   &   -0.0068\%                             \\ \bottomrule
\end{tabular}
\caption{The results of a few measures, calculated and averaged across a 5-fold cross-validation setup. The receipt forecasts are compared against the actual receipts within the $k^\text{th}$ subset per technique, having trained the technique on the rest of the data. The PAR-metric expresses the sum of discounted shortfalls (essentially `arrears') between instalments and forecasts as a proportion of all gross advances, using 7\% as the discounting rate. The actual PAR-value is -1.64\% on average, which is negatively signed due to large historical overpayments at earlier periods.} \label{tab:validation_tests}
\end{table}

We experimented with a few measures that span forecast error, portfolio impact, and overall parameter stability, with some of the results thereof given in \autoref{tab:validation_tests}. One of these measures is the PAR-metric, which reflects the portfolio-wide arrears rate as implied when replacing historical receipts with each technique's forecasts. The PAR of the Markovian technique ($T_b$) is much closer to the actual rate than that of the simpler technique ($T_a$), which attests to at least two advantages of the former. $T_b$ incorporates the possibility of curing and produces forecasts based on the level of accrued delinquency, both of which are ignored by $T_b$. The forecast accuracy was initially assessed on the account-level as well (using Mean Absolute Error), though was later deemed too onerous a measure for our particular context. Regarding parameter stability, the mean \%-difference in parameter estimates is reassuringly close to 0, as calculated between using all data versus using each training fold in our cross-validation setup.

\section{Loss-optimising recovery decision times: an empirical illustration}
\label{sec:5}

The parameters of each forecasting technique were previously estimated from three progressively worse samples, thereby recognising that a portfolio's historical risk composition itself will bias the forecast receipts. Naturally, the LROD-procedure itself can be applied once onto each of these subsequent samples. The recovery threshold-locations at which losses are minimised (if found) are expected to differ significantly, given the different risk profiles. That said, this procedure is imagined to be applied on the entire loan portfolio when loss-optimising a bank's recovery decision in practice. It is, however, iteratively applied in this study as an experimental and artificial proxy for various risk compositions found in reality, as if each sample is a stand-alone portfolio. Moreover, the sample $S_i$ from which a forecasting technique is parametrised (or trained) may differ from the sample $S_j$ on which the LROD-procedure is applied, where both $i$ and $j$ are indexes that denote samples $\{S_1,S_2,S_3\}$. Apart from using data more efficiently, this approach approximates the reality of a portfolio's historical risk composition changing in the future. As an example, training a forecasting technique on $S_3$ but optimising recovery thresholds on $S_1$ simulates the context of a proportionally lower-risk portfolio ($S_1$) undergoing heavy financial strain in the future (by using forecasts trained from $S_3$). Additionally, this proposed setup aligns with the IFRS 9 accounting standard, which requires expected credit losses to be estimated based on various macroeconomic scenarios, as stated in \citet[par.~5.5]{ifrs9_2014}. Therefore, the experimental setup is illustrated as a $3\times3$ matrix in \autoref{tab:setup} wherein each cell $s_{ij}$ represents the results from a specific scenario. Greater values of $j$ denote riskier portfolios, while greater values of $i$ represent more pessimistic forecasts.

\begin{table}[ht!]
\centering
\begin{tabular}{@{}cllll@{}}
 \multicolumn{1}{l}{} &  & \multicolumn{3}{c}{$j$} \\ 
 \multicolumn{1}{l}{} &  & $S_1$ & $S_2$ & $S_3$ \\ \cline{3-5} 
 \multirow{3}{*}{$i$} & \multicolumn{1}{c|}{$S_1$} & $s_{11}$    & $s_{12}$    & \multicolumn{1}{l|}{$s_{13}$} \\
 & \multicolumn{1}{c|}{$S_2$} & $s_{21}$    & $s_{22}$    & \multicolumn{1}{l|}{$s_{23}$} \\
 & \multicolumn{1}{c|}{$S_3$} & $s_{31}$    & $s_{32}$    & \multicolumn{1}{l|}{$s_{33}$} \\ \cline{3-5} 
\end{tabular}
\caption{The experimental setup containing nine scenarios wherein row $i$ represents the sample used for parametrising a forecasting technique, and column $j$ denotes the sample on which optimisation is performed.} \label{tab:setup}
\end{table}

On interpreting the following results, the LROD-procedure's particular loss model assumes that a portion of the expected balance and arrears amount are immediately lost when entering $(g,d)$-default. In effect, this equates the actual default and write-off events to a single point, which implies that a loss-optimised threshold is not necessarily the supposed starting point of legal proceedings, but rather the optimal ending point thereof. For that matter, finding the optimal starting point suggests that the length of the subsequent workout period will vary. Other than data challenges, there are a few operational and legal factors that may influence the workout length, which could require a more comprehensive loss model than the one used in this study. Therefore, finding the best starting point of legal proceedings is left as an avenue of future investigation.

\subsection{Optimisation results using \texorpdfstring{$S_1$}{Lg}, \texorpdfstring{$S_2$}{Lg}, and \texorpdfstring{$S_3$}{Lg} respectively}
\label{sec:5-Results}

\begin{figure}[ht!]
\centering\includegraphics[width=0.8\linewidth,height=0.5\textheight]{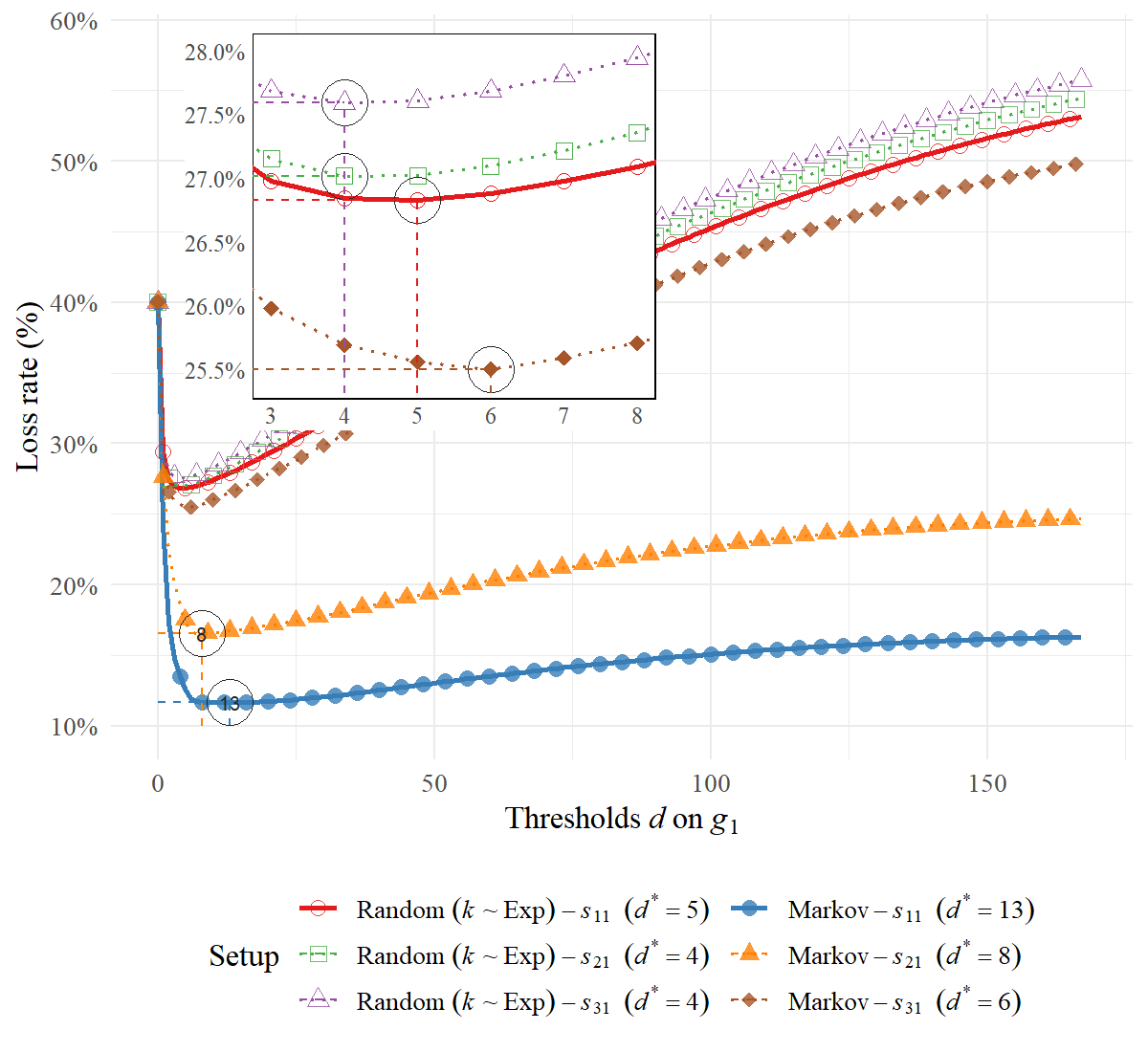}
\caption{Loss rates across recovery thresholds $d$ for measure $g_1$ on the full sample $S_1$ across various forecasting scenarios, using the random defaults technique with $k\sim\text{Exp}(\lambda)$ truncation, and using the Markovian defaults technique independently. Solid lines indicate a base scenario wherein both optimisation and training forecasts use the same sample. Zoomed plot and encircled points show global minima for each loss curve, also bracketed in the legend.}\label{fig:LossThresh_s1}
\end{figure}

The first set of results are presented in \autoref{fig:LossThresh_s1} wherein all loss curves exhibit minima at certain thresholds $d^\ast$ when loss-optimising the recovery decision on the full sample $S_1$, i.e., results based on the first column in \autoref{tab:setup}. Optimising on $S_1$ represents a historically lower-risk portfolio, while training forecasts from $\{S_1,S_2,S_3\}$ represents increasingly dire credit risk scenarios in future. Specifically, the loss minimum increases both in value as well as occur at decreasing thresholds as the forecast scenario worsens, i.e., progressing from $s_{11}\rightarrow s_{21}\rightarrow s_{31}$ when parametrising the forecasting technique. Furthermore, overall losses across all thresholds increase as the forecast scenario deteriorates, which is evidenced by the steeper slope of the loss curve after having reached its minimum at $d^\ast$. This agrees with the intuition of cutting losses sooner rather than later when facing increasingly higher credit risk on future cash flows. Moreover, the $s_{31}$-results yielded the lowest thresholds $d^\ast = \{4,6\}$ respective to each technique, whose values seem close to the current practice of using $d=3$ with the $g_0$-measure as a default definition. Therefore, training forecast models from $S_3$ may serve as a conservative `boundary' case, thereby deliberately introducing risk aversion when optimising the recovery decision itself. 

\begin{figure}[ht!]
\centering\includegraphics[width=0.8\linewidth,height=0.5\textheight]{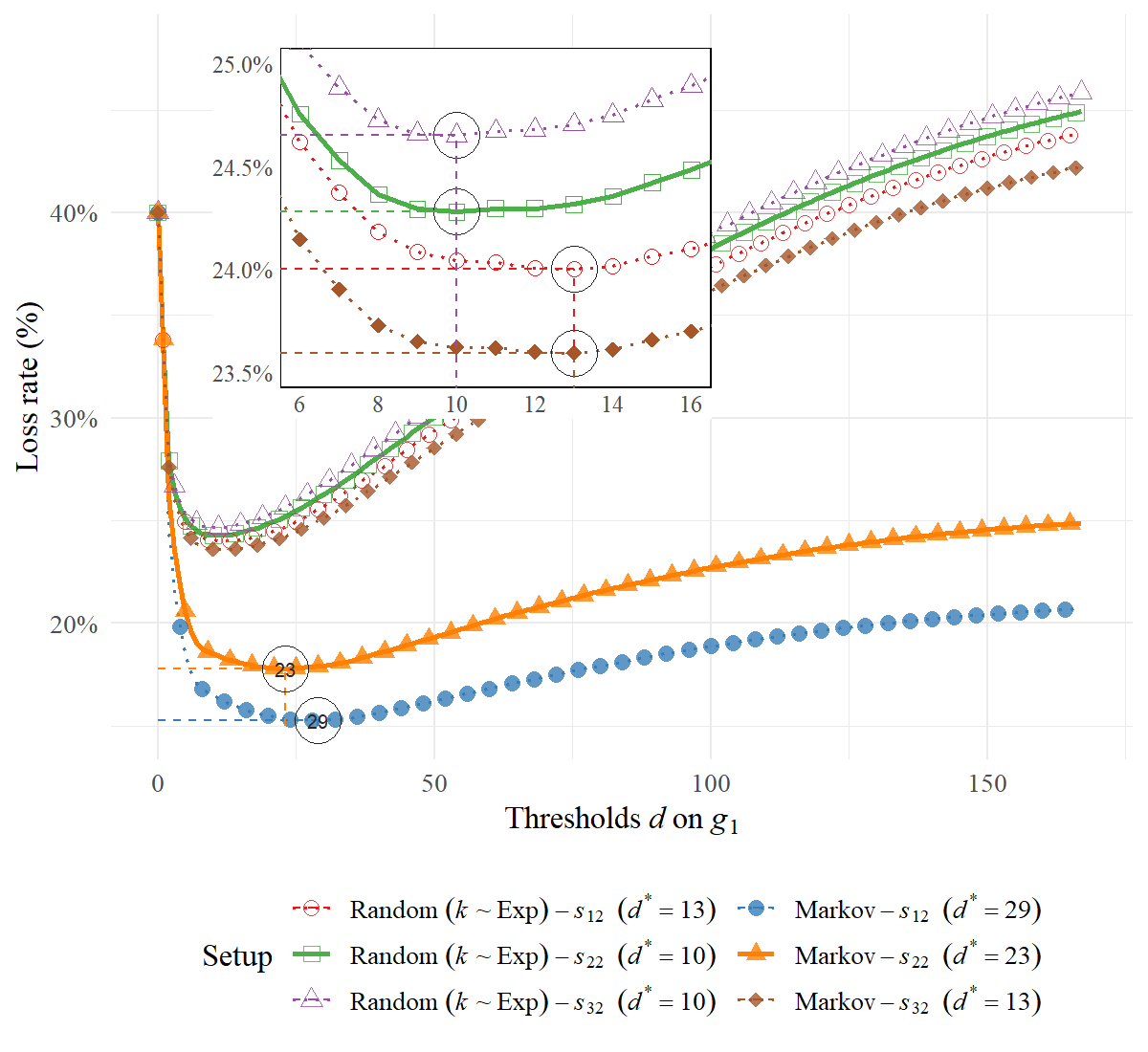}
\caption{Loss rates across recovery thresholds $d$ for measure $g_1$ on the delinquents sample $S_2$ across various forecasting scenarios, using the random defaults technique with $k\sim\text{Exp}(\lambda)$ truncation, and using the Markovian defaults technique independently. Graphical formatting follows that of \autoref{fig:LossThresh_s1}.}\label{fig:LossThresh_s2}
\end{figure}

Regarding the techniques, the base scenario $s_{11}$ clearly gives two very different loss minima at $d^\ast=5$ for random defaults versus $d^\ast=13$ for Markovian defaults. Incidentally, the latter also yielded lower loss rates at less stringent (higher) values of $d\ast$ in general, when compared to those given by random defaults across all scenarios $s_{i1}$. Moreover, the difference between $d^\ast$ yielded by each technique becomes smaller as the forecast scenario worsens. In fact, Markovian defaults for $s_{31}$ gives a loss curve that is similarly shaped to those produced by random defaults irrespective of forecast scenario, which suggests some connection. Consider that the underlying transition matrix in \autoref{tab:markov2} is generally much more transient than the one in \autoref{tab:markov1}, with far greater conditional probabilities of transiting to worse states. As the possibility of curing back to a better state declines, the Markovian technique increasingly resembles the simpler technique in effect. The latter provides inherently less realistic forecasts since it deliberately ignores curing, which means the larger cash flows associated with curing events are not generated. Since the Markovian forecasts are demonstrably more accurate, they are therefore clearly preferable, though the random forecasts are kept for expositional purposes, including model risk.

\begin{figure}[ht!]
\centering\includegraphics[width=0.8\linewidth,height=0.5\textheight]{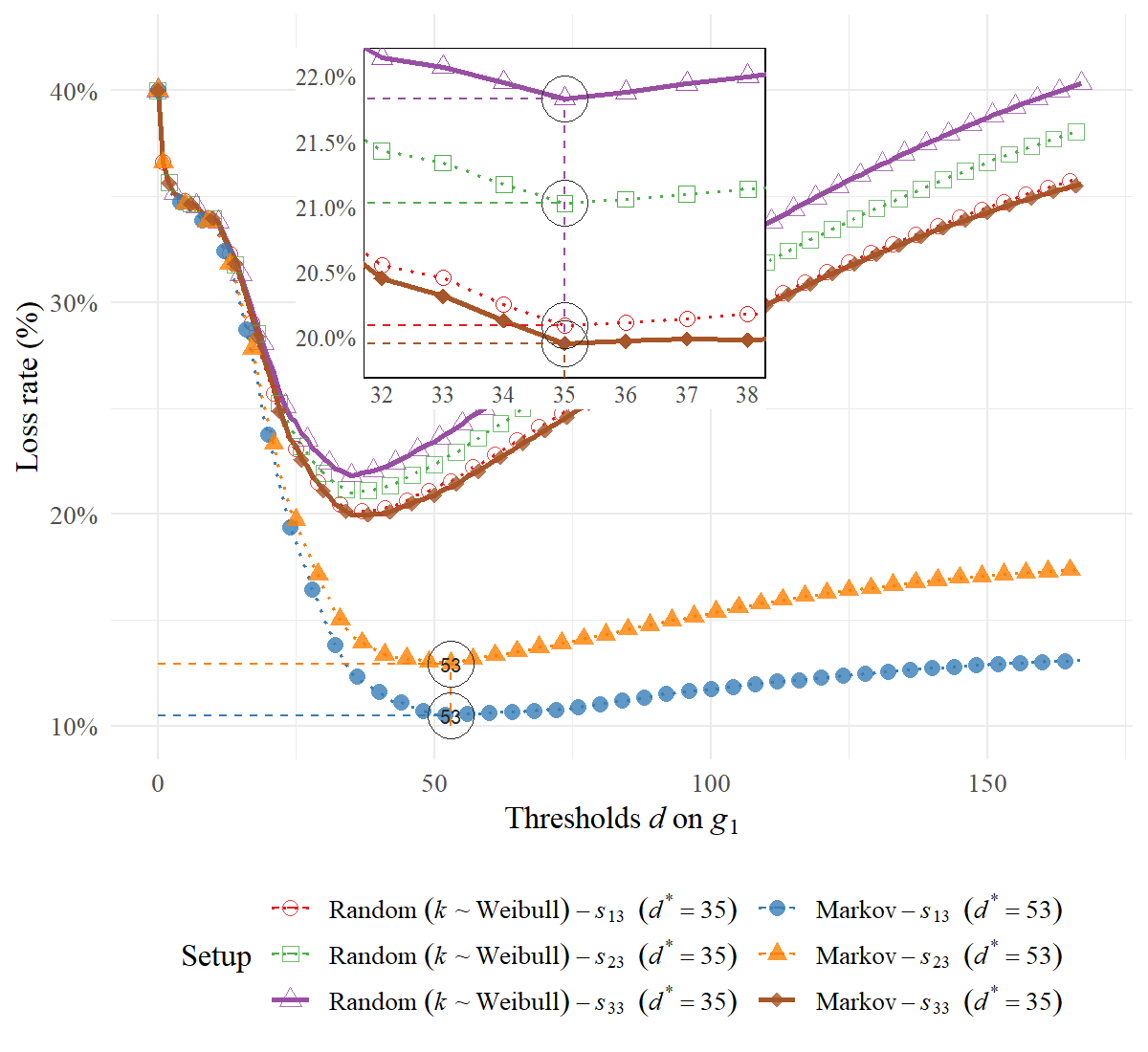}
\caption{Loss rates across recovery thresholds $d$ for measure $g_1$ on the write-offs sample $S_3$ across various forecasting scenarios, using the random defaults technique with $k\sim\text{Weibull}(\lambda,\phi)$ truncation, and using the Markovian defaults technique independently. Graphical formatting follows that of \autoref{fig:LossThresh_s1}.}\label{fig:LossThresh_s3}
\end{figure}

The same trends hold true in \crefrange{fig:LossThresh_s2}{fig:LossThresh_s3} when optimising on samples $S_2$ and $S_3$ instead, i.e., scenarios from the second/third columns in \autoref{tab:setup}. As the main result, optima are still obtained (though at different locations) across all techniques and forecast scenarios, thereby demonstrating the LROD-procedure's sensitivity to the inherent risk profile of a portfolio. In particular, optimising across increasingly riskier portfolios ($S_1\rightarrow S_2 \rightarrow S_3$) remains viable, even if the loss curves become somewhat vertically compressed relative to lower-risk samples. Moreover, $d^\ast$ seem to \textit{increase} across riskier samples. The base scenarios $\{s_{11},s_{22},s_{33}\}$, i.e., the diagonal in \autoref{tab:setup}, demonstrate this phenomenon with loss minima found at $d^\ast = \{5,10,35\}$ for random defaults and $d^\ast = \{13,23,35\}$ for Markovian defaults. That said, the recoveries realised from selling the underlying asset largely explains this phenomenon. Since these recoveries are generally recognised only at the write-off point after a typically long workout period, the suddenly large receipt will dramatically decrease the delinquency level at the last period. Therefore, when optimising on $S_3$, it is indeed statistically better to wait strategically and collect some of these large cash flows. This is evidenced by the relatively high threshold $d^\ast=35$, which indicates the optimal \textit{ending} point of legal proceedings. The fact that both loss minima and their thresholds change when optimising on $S_3\rightarrow S_2 \rightarrow S_1$ merely attests to the dilution of written-off cases as a proportion of the overall portfolio.

While the \textit{CD}-measure $g_1$ is primarily used in this study, loss-optimality was also found for the other two measures $g_2$ and $g_3$ across the experimental setup for both forecasting techniques. Interestingly, the $d^\ast$ yielded by $g_2$ and $g_3$ are much less varied than those yielded by $g_1$, which suggests these measures are not as sensitive as $g_1$ to the choice of technique, risk level, or forecast scenario. Specifically, the loss minima for $g_2$ and $g_3$ occur within the threshold ranges $[1.2,1.9]$, $[1.3,2.3]$, and $[3.2,6]$ when optimising respectively across $\{S_1,S_2,S_3\}$. However, the loss minima themselves are greater than those yielded by $g_1$ with the percentage difference thereof averaging at $3.6\%$. Evidently, the LROD-procedure suggests that the $g_1$ measure is objectively the best delinquency measure for signalling loan recovery -- at least for this particular mortgage portfolio.

\subsection{A Monte Carlo-based refinement for analysing the variance of optima}
\label{sec:5-MonteCarlo}

Given that forecast receipts are inherently probabilistic, their subsequent use within a delinquency measure injects uncertainty into the latter's output as well as into the optimisation itself. Therefore, any loss minimum that is found at a certain threshold may, in fact, be spurious. As an example, a random but systemic perturbation at some time point in the underlying forecasts can produce an alternative minimum at an entirely different threshold, which has implications for the overall precision of the optimisation. Confidence in this supposed minimum can be enhanced by conducting a variance study of sorts on the loss curve. One approach to this problem is to produce multiple sets of forecasts of the portfolio's future cash flows using simple Monte Carlo simulation and the laws of large numbers. Each iteration thereof will have its own independent loss curve using a particular set of random forecasts generated from a specific technique. As an example, consider $n$ such Monte Carlo trials, thereby resulting in $n$ loss rate estimates at each threshold $d$, from which a sample mean $\mu_d$ is calculated at each $d$. The corresponding sample variance $s^2_d$ is estimated, which is finally used in constructing a standard 99\% confidence interval for the mean as $\mu_d \pm 2.58 \, s_d/\sqrt{n}$.

\begin{figure}[ht!]
\centering\includegraphics[width=0.8\linewidth,height=0.5\textheight]{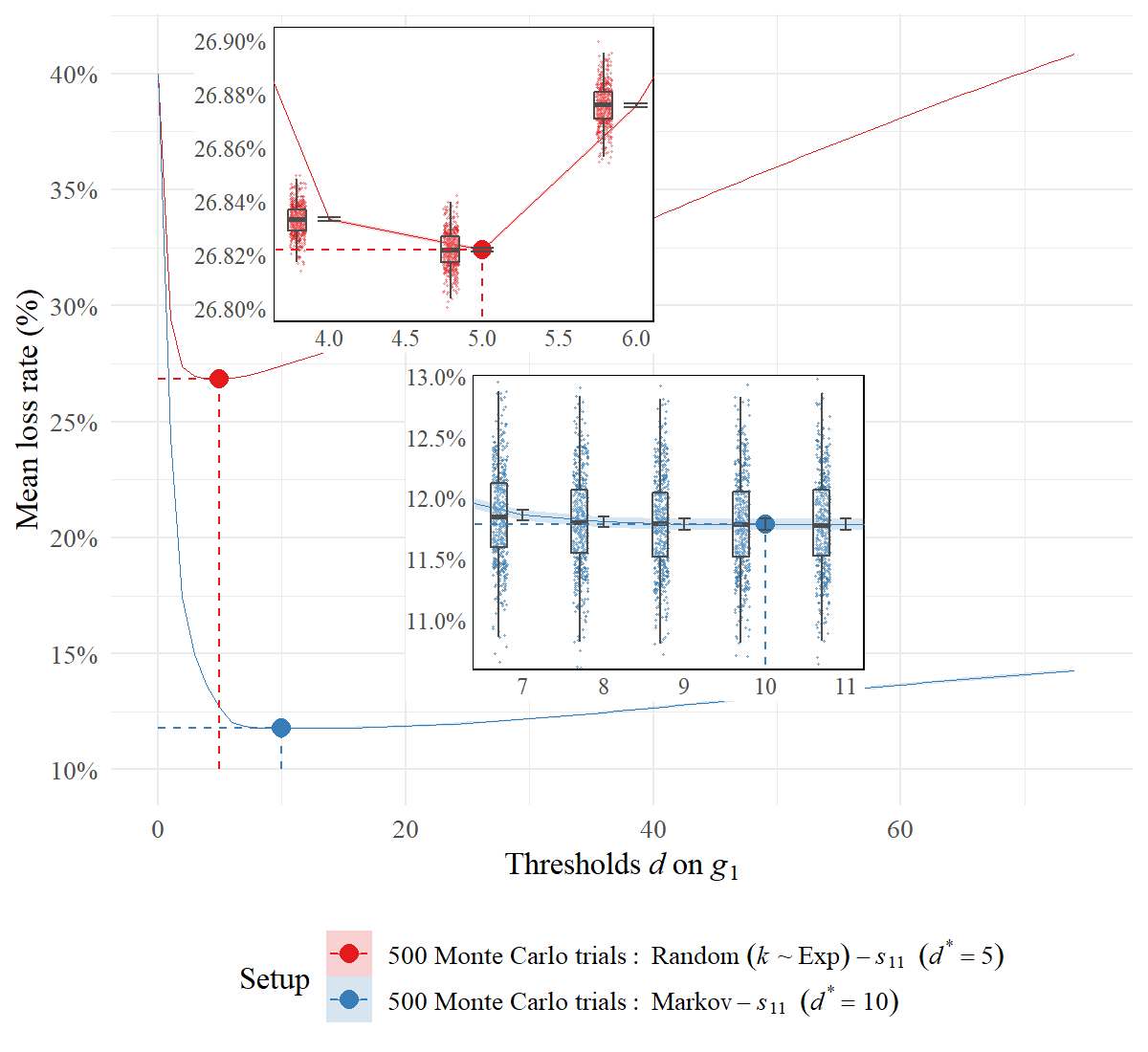}
\caption{Average loss rates (solid lines) across thresholds $d$ for measure $g_1$, estimated from $n=500$ Monte Carlo trials, for scenario $s_{11}$. Forecasts are iteratively and independently made using the random defaults technique with $k\sim\text{Exp}(\lambda)$ truncation, and the Markovian defaults technique. The averages are accompanied by a 99\% shaded confidence band with error bars. Zoomed plots show global minima for each loss curve, also bracketed in the legend. Box-and-whiskers mini-plots within the zoomed plots summarise the overlaid loss estimates at each $d$.}\label{fig:LossThresh_s1_MC}
\end{figure}

Monte Carlo simulation is illustrated in \autoref{fig:LossThresh_s1_MC} for both forecasting techniques using the $s_{11}$ base scenario after 500 runs. The forecasts yielded by the simpler technique appear to be quite robust since the resulting loss rates had relatively little variation and retained the overall shape of the original loss curve in \autoref{fig:LossThresh_s1}. These results, particularly the difference in the widths of the confidence intervals per technique, attest to the \textit{bias-variance} trade-off phenomenon in statistical learning, as the model's complexity varies. More specifically, the simpler technique with its relatively invariant forecasts is also much less accurate than the Markovian technique. Reassuringly, the lowest sample mean still occurs at $d^\ast=5$ as it did previously. However, the same cannot be said for the Markovian forecasts since the minimum now occurs at $d^\ast=10$ (down from the previous $d^\ast=13$). While the overall shape of the Markovian loss curve is still the same, the loss rate estimates exhibit greater variance than those of the simpler technique. Moreover, the loss curve is relatively flat in the region near $d^\ast=10$ (as in \autoref{fig:LossThresh_s1}), which helps explains the `ease' at which the minimum shifted in this particular scenario.

\begin{figure}[ht!]
\centering \vspace{-10pt} 
\begin{subfigure}{0.67\textwidth}
    \caption{Scenario $s_{22}$ (delinquents)}
    \centering\includegraphics[width=1\linewidth,height=0.43\textheight]{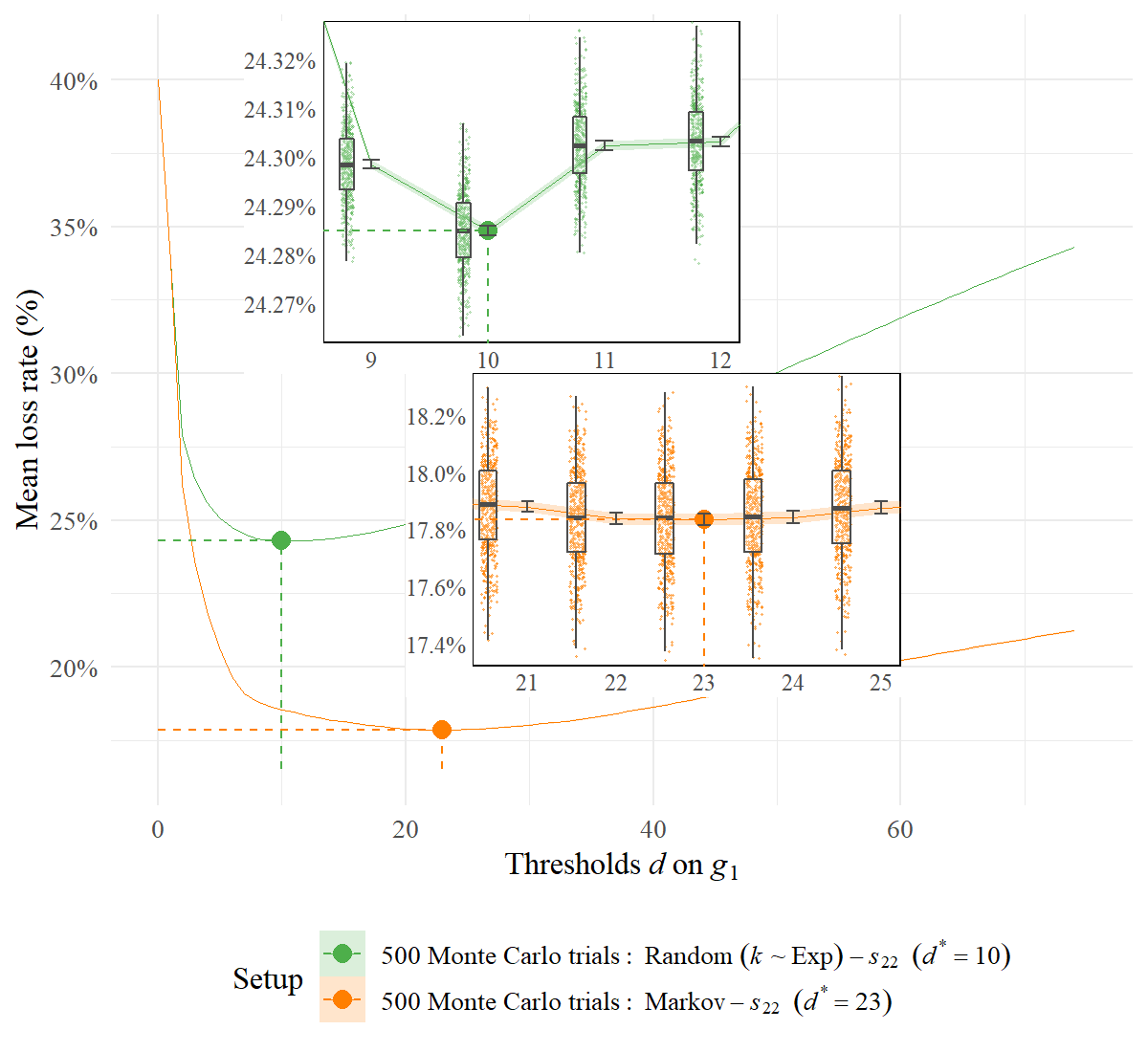}\label{fig:LossThresh_s2_MC}
\end{subfigure} \\ 
\begin{subfigure}{0.67\textwidth}
    \caption{Scenario $s_{33}$ (write-offs)}
    \centering\includegraphics[width=1\linewidth,height=0.43\textheight]{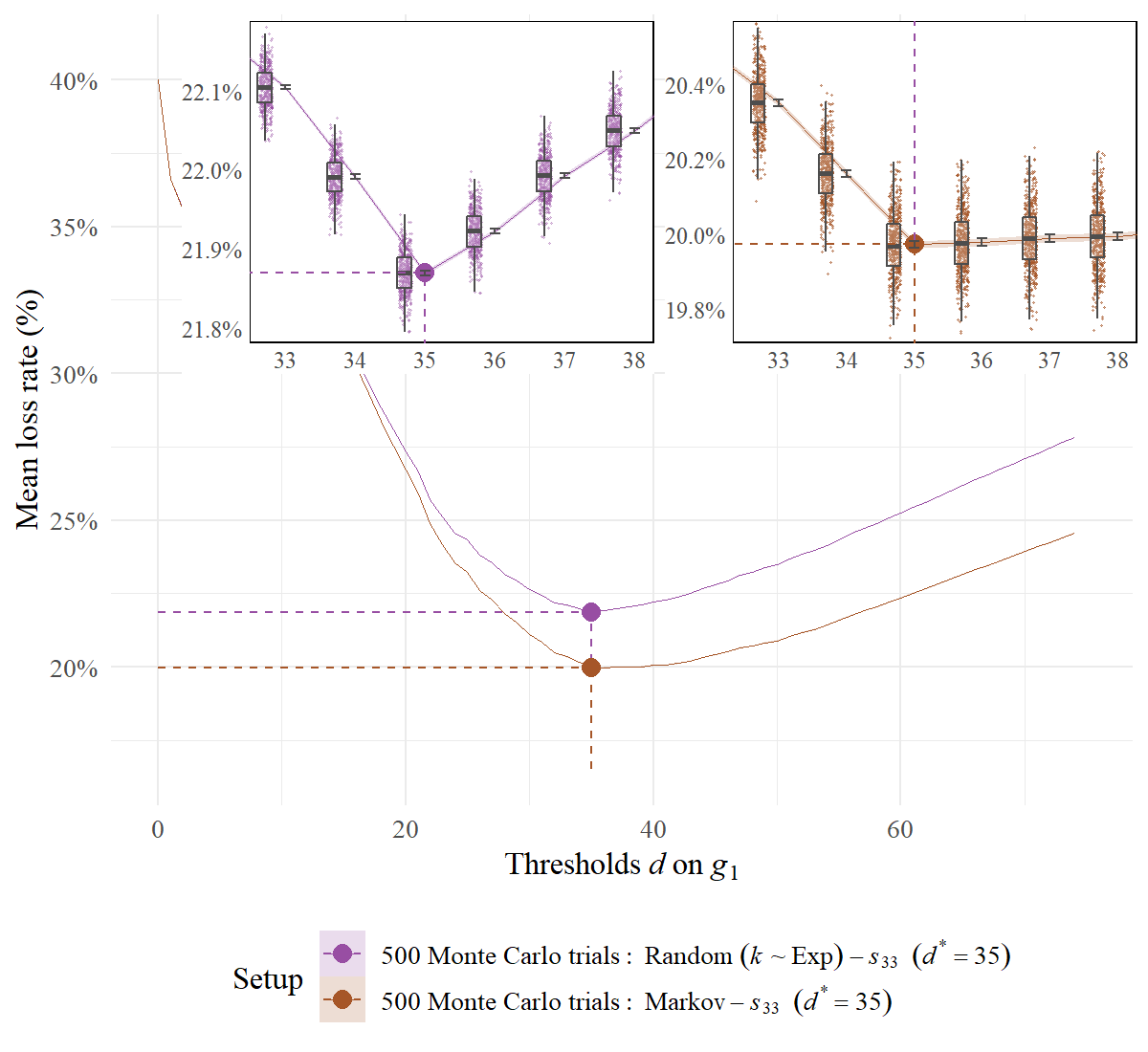}\label{fig:LossThresh_s3_MC}
\end{subfigure} \vspace{-15pt}
\caption{Average loss rates (solid lines) across thresholds $d$ for measure $g_1$, estimated from $n=500$ Monte Carlo trials, for scenarios $s_22$ and $s_33$. Forecasts are iteratively and independently made using the random defaults technique with $k\sim\text{Exp}(\lambda)$ truncation in \textbf{(a)} and with $k\sim\text{Weibull}(\lambda,\phi)$ truncation in \textbf{(b)}, with Markovian forecasts provided in both panels. Graphical formatting follows that of \autoref{fig:LossThresh_s1_MC}.}\label{fig:LossThresh_s2_s3_MC}
\end{figure}

Conducting these Monte Carlo simulations clearly refines the LROD-procedure one step further by controlling for the uncertainty within forecasts. That said, it is not necessarily true that the average minimum loss will \textit{always} occur at a different threshold, as it did in \autoref{fig:LossThresh_s1_MC}. In fact, the average minima remained at the same thresholds as they did in \crefrange{fig:LossThresh_s2}{fig:LossThresh_s3}, when running these Monte Carlo simulations for the other base scenarios $s_{22}$ and $s_{33}$ in \autoref{fig:LossThresh_s2_s3_MC}, regardless of forecasting technique. Moreover, the general shape of each loss curve \autoref{fig:LossThresh_s2_s3_MC} remained the same, all of which provides combined assurance on the precision of the optimisation results. Lastly, the practitioner may consider a smaller and more focused range of thresholds, especially within the general region of optima, when conducting these Monte Carlo simulations in practice. In contrast, we chose a larger range to demonstrate the LROD-procedure (and its viability) in a principled manner as a "proof of concept".

\section{Conclusion}
\label{sec:6}

The timing of the recovery/foreclosure decision is empirically illustrated as a delinquency-based optimisation problem, such that loans are forsaken neither too early nor too late, if at all. The present study refines and extends our previous simulation-based work in \citet{botha2020paper1}, though using a real-world mortgage portfolio from the South African market. However, real-world datasets are often right-censored such that many accounts have not yet reached contractual maturity, which poses an additional and non-trivial challenge. In fact, experimentation shows that \textit{not} treating a portfolio for censoring leads to uninteresting and unintuitive results during optimisation. While an uncensored portfolio would be ideal, the paucity of both data and lenders willing to avail sufficiently rich data makes this difficult. Moreover, most portfolios are actively grown by banks, which causes right-censoring and implies that recovery optimisation will likely remain problematic in practice. We demonstrate a more feasible remedy wherein available data is first used to forecast the residual cash flows of each account up to its contractual maturity. This step `completes' the portfolio and enables the practical use of the LROD-procedure for optimising the bank's recovery decision.

As a secondary contribution, two forecasting techniques are proposed, parametrised, and applied on the portfolio before optimisation. This includes a simple probabilistic technique and a more sophisticated eight-state Markov chain, both of which are subsequently used in forecasting cash flows independently. However, the manner in which receipts are forecast will greatly affect the portfolio's subsequent credit risk profile, which influences the timing of loan recovery at which the minimum loss is subsequently attained. Accordingly, forecasts are artificially differentiated by training them from different account subsets (or samples), where each sample contains a progressively greater proportion of delinquent accounts by design. Effectively, each sample approximates a different risk composition typically found in reality, e.g., mortgages vs. unsecured loans, as if each sample is a stand-alone portfolio. Furthermore, the sample on which we optimise may differ from the sample from which forecasts are trained. This simulates the reality of a portfolio's historical risk composition changing in the future by forecasting receipts accordingly, while also making more efficient use of data. Additionally, this setup aligns with IFRS 9 by using various macroeconomic scenarios when estimating expected losses.

Within each scenario of this experimental setup, we find a so-called `Goldilocks'-region that contains an \textit{ideal} delinquency threshold at which the portfolio loss is minimised. This setup demonstrates that the LROD-procedure is sensitive to the historical risk profile of a portfolio. Moreover, riskier forecasts yield smaller (or more stringent) optimal delinquency thresholds, which agrees intuitively with cutting losses sooner rather than later as risk expectations deteriorate. In addition, we contribute a Monte Carlo-based procedural refinement that can provide additional assurance on the stability of optima, especially given the uncertainty underlying all forecasts. To this point, the choice of forecasting technique itself affects recovery optimisation, which is demonstrated by the significant differences between each technique's optima. By design, the Markovian technique is much more realistic since it allows for curing backwards to lower delinquency levels, which affects the size of forecasts. Conducting a 5-fold cross-validation further verified the superior quality of Markovian forecasts. However, the simpler technique is retained for expositional purposes since it clearly demonstrates the dangers of model risk when forecasting. That said, an ensemble of forecasting techniques suggests that a meta-learning approach may be viable, which can certainly be further examined in future work. For example, optima can be averaged across technique and forecast scenario using a weighting scheme of sorts. Besides optimisation, there is also practical value for developing these forecast models within an IFRS 9-compliant loss provisioning context. 

Regarding limitations, historical cash flows are surely affected by past collection strategies (and their subsequent success or failure) that were employed by the bank at the time. Therefore, training a forecast model from the same data carries the unavoidable risk of embedding the effects of previous strategies into the optimisation, as additional data `noise'. Future research can perhaps focus on controlling for the bank's strategic influence on these cash flows over time when forecasting receipts. Another avenue of future study is to explore a finer-grained segmentation scheme during the optimisation step. Partitioning data into three increasingly riskier samples (as we did) correctly assumes homogeneity within each sample. However, recovery decision times can surely be further optimised within certain segments of the portfolio, instead of yielding a portfolio-wide criterion. This may attenuate the LROD-procedure further to the idiosyncrasies of a portfolio, though one will have balance greater segmentation against too little data within a segment. Furthermore, future work can certainly explore a less censored (and therefore richer) portfolio of shorter-term loans, which can reduce the necessary forecasting extent as well as improve the forecasting ability. Lastly, future studies can expand upon the current loss model by incorporating dynamic cost components more explicitly, e.g., funding costs. The static loss rates $r_E$ and $r_A$ may be converted into proper LGD-models instead such that loss rates are estimated from the time of entering our particular $(g,d)$-default state. Pursuing this particular avenue will likely intersect with the literature on credit risk modelling and IFRS 9, which can enhance model sophistication given that the field itself is currently in vogue.

%--------------------------------------------------------%
%	REFERENCE LIST
%--------------------------------------------------------%

%TC:ignore

\singlespacing
\printbibliography % using biblatex
%\section*{References}
%\bibliographystyle{newapa}
% see http://texdoc.net/texmf-dist/doc/latex/natbib/natbib.pdf for more styles
%\bibliography{bibliography} 
\onehalfspacing

\section*{Appendix}
\label{app:app01}

\subsection{Failing to forecast before recovery time optimisation}
\label{sec:untreated_portfolio}

\begin{figure}[ht!]
\centering\includegraphics[width=0.8\linewidth,height=0.5\textheight]{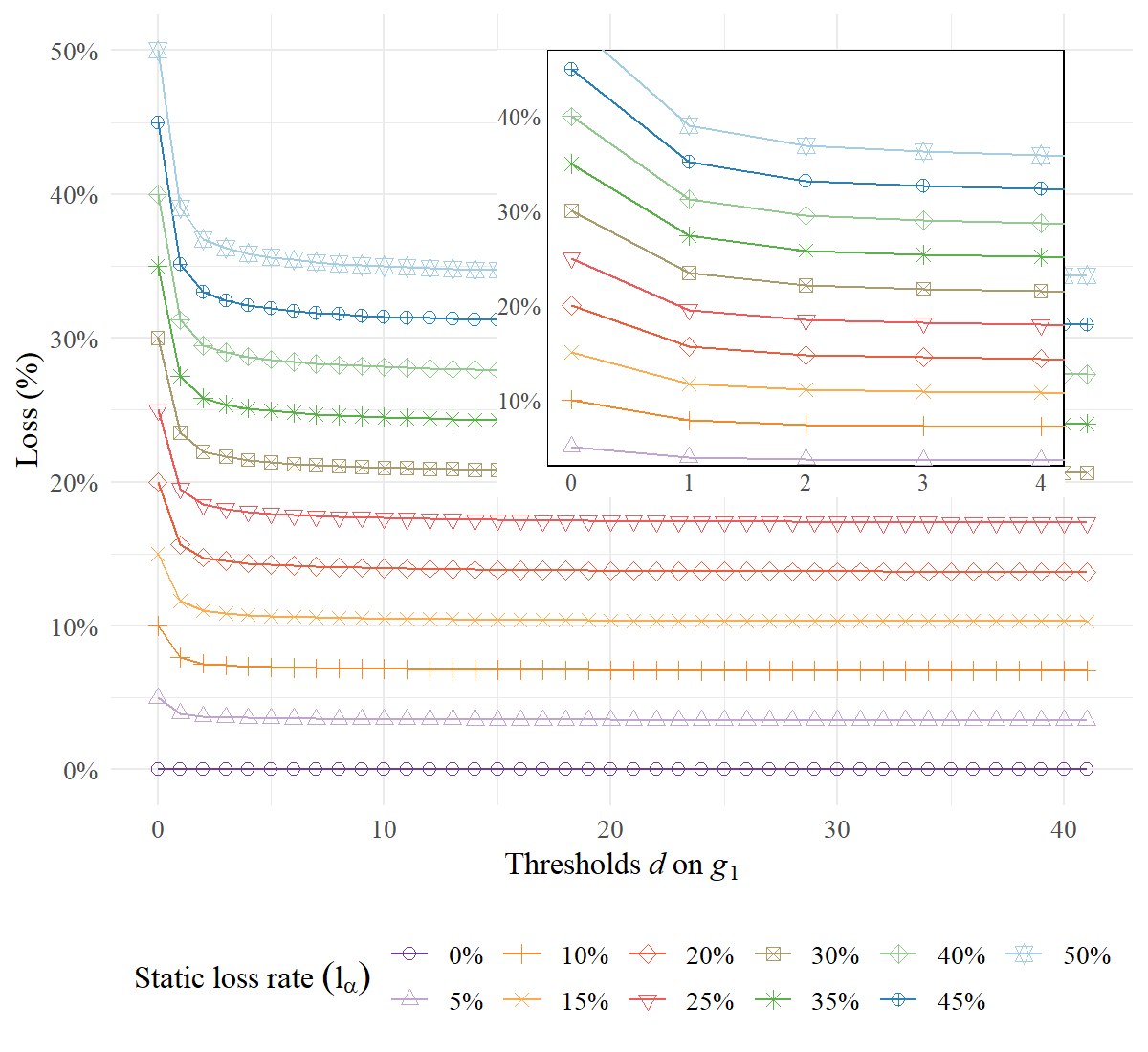}
\caption{Losses (expressed as a \% of the summed principals) across default thresholds $d$ for the \textit{CD}-measure $g_1$, using a range of static loss rates $l_\alpha \in [0,1]$ and an untreated real-world loan portfolio.}\label{fig:Untreated_LossThresh}
\end{figure}

A real-world loan portfolio typically has immature accounts, i.e., the receipt vector $\boldsymbol{R}=\left[R_1,R_2,\dots,R_{t_0}\right]$ of an immature account only contains observed elements from data up to the most recent time point $t_0 < t_c$. The remaining future elements $t_1,\dots,t_c$ are unobservable and deliberately ignored in this exercise to demonstrate the absence of forecasting in the results of the LROD-procedure. Furthermore, the balances of each account are observed at relevant time periods and simply multiplied with a static loss rate $l_\alpha \in [0,1]$, as a simpler loss model. More specifically, the most recent balance at time $t_0$ is used for a $(g,d)$-performing account whilst the balance at the default time $\tau \leq t_0$ is used for a $(g,d)$-defaulting account, as signalled by a particular $(g,d)$-configuration. In both cases, the observed balance is simply discounted back to time $t=1$ (loan origination) using the same 7\% risk-free rate. Selecting a range of loss rates at will, the LROD-procedure is then iteratively applied on the entire portfolio. The resulting loss curves are presented in \autoref{fig:Untreated_LossThresh} using the \textit{CD}-measure $g_1$. There are no significant differences in the shapes of loss curves for loss rates exceeding 50\%.

No global minima in losses exist at any particular threshold, regardless of the chosen loss rate. Instead, all losses tend toward a certain asymptote that is influenced by the loss rate, which renders the optimisation of the recovery threshold a moot point. Moreover, the resulting loss curves suggest that one should simply ignore any accrued delinquency, except for very low thresholds $d\leq2$, which coincide with the greatest losses. Although unusual, consider that the majority of the portfolio's receipts are still pending. The LROD-procedure's particular loss model recognises this and logically suggests never to recover a single account. Regardless, ignoring accrued delinquency at large is intuitively false and ill-advised for a credit risk-based business like a bank. Instead, this result rather attests to the breakdown of the LROD-procedure itself when foregoing the necessary forecasting of a loan portfolio's cash flows.

\subsection{Fitting statistical distributions to the truncation parameter \texorpdfstring{$k$}{Lg}}
\label{sec:StatFit}

In calibrating the random defaults technique to forecast cash flows, a truncation effect simulates the reality that some accounts will simply never resume payment. This is achieved when the forecast receipts are zero-valued after a certain point $t_k$ coinciding with reaching a certain $k$-threshold in measured delinquency. In estimating this $k$ truncation parameter, the maximum observed delinquency (using $g_1$) per account is calculated with the resulting empirical distributions of these maxima shown in \autoref{fig:Dist_Max_CD}, respective to each sample $S_2$ (delinquents) and $S_3$ (write-offs). Several candidate statistical distributions are then fit using maximum likelihood on each respective sample, with the fitted probability density function overlaid on the histogram, as shown in \autoref{fig:StatFit} for some of these candidates.

In selecting the best fit, both Kolmogorov-Smirnov and Anderson-Darling goodness-of-fit tests are conducted for each candidate distribution against the standard 5\% significance level. However, all of the null hypotheses are rejected for both $S_2$ and $S_3$, presumably due to the heavily right-skewed distributions of maxima in both cases. Secondly, the Akaike Information Criterion (AIC) reveals that the Dagum, log-normal, Pareto, Weibull, exponential, and gamma distributions were amongst the best fitting candidates for $S_2$. Though the Dagum distribution had the best AIC, the exponential distribution is chosen owing to its simplicity and its somewhat greater popularity in statistical literature. Furthermore, the exponential distribution is strictly decreasing for $x$, which is deemed more appropriate given the histogram's shape. Similarly, the AIC for $S_3$ suggests that the Dagum, Burr (Type 12), Weibull, Gumbel, Gamma, and Logistic distributions were the better-fitting candidates. Of these, the Weibull distribution is chosen since it best approximates the histogram visually without lending too much credence to the left-tail though still yielding a sufficiently heavy right-tail.

\afterpage{%
\clearpage 
\thispagestyle{empty}
\begin{figure}[ht!]
\centering \vspace{-21pt} 
\begin{subfigure}{0.75\textwidth}
    \caption{Using the delinquents sample $S_2$}
    \centering\includegraphics[width=1\linewidth,height=0.46\textheight]{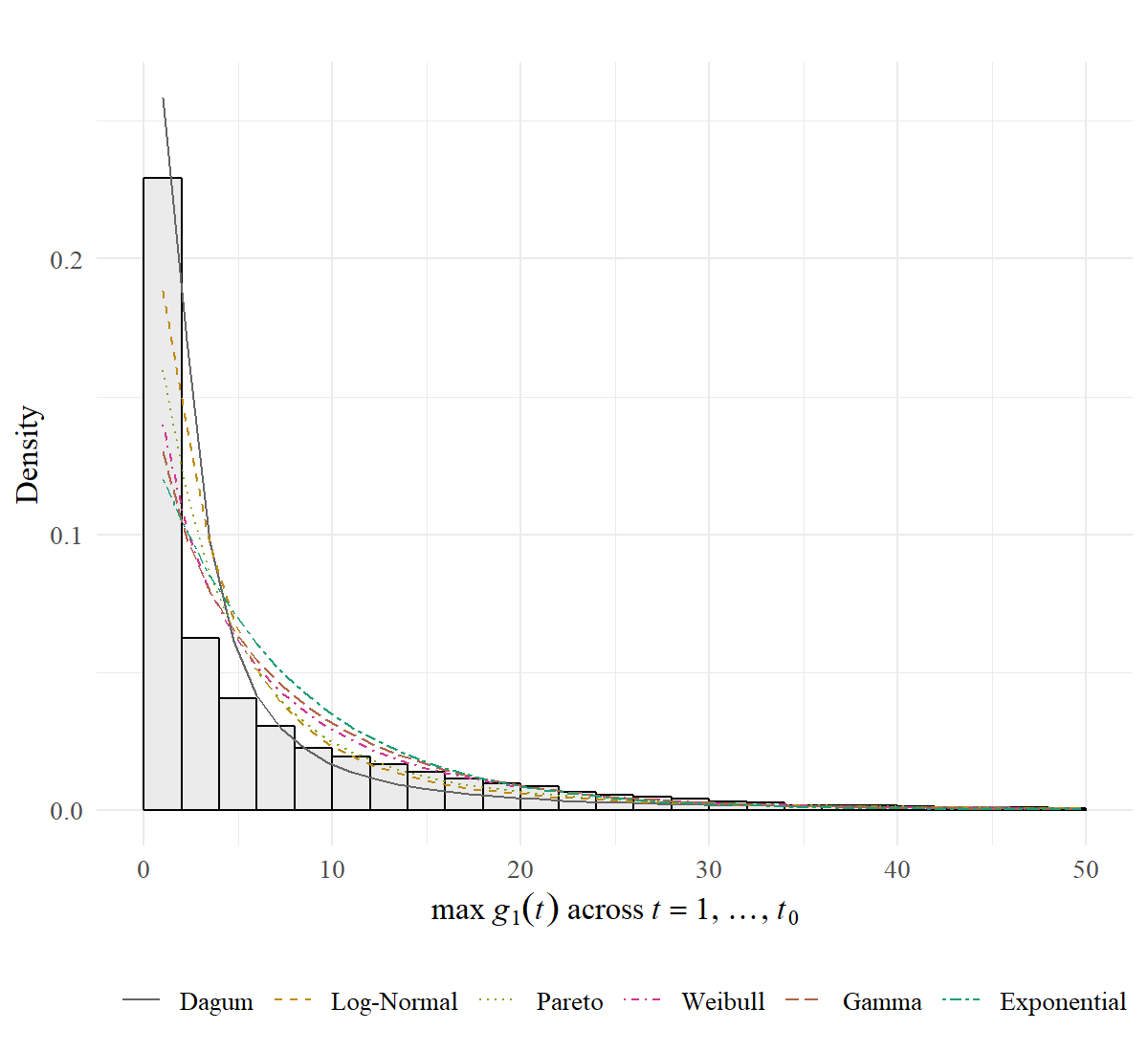}\label{fig:StatFit_a}
\end{subfigure} \\ 
\begin{subfigure}{0.75\textwidth}
    \caption{Using the write-offs sample $S_3$}
    \centering\includegraphics[width=1\linewidth,height=0.46\textheight]{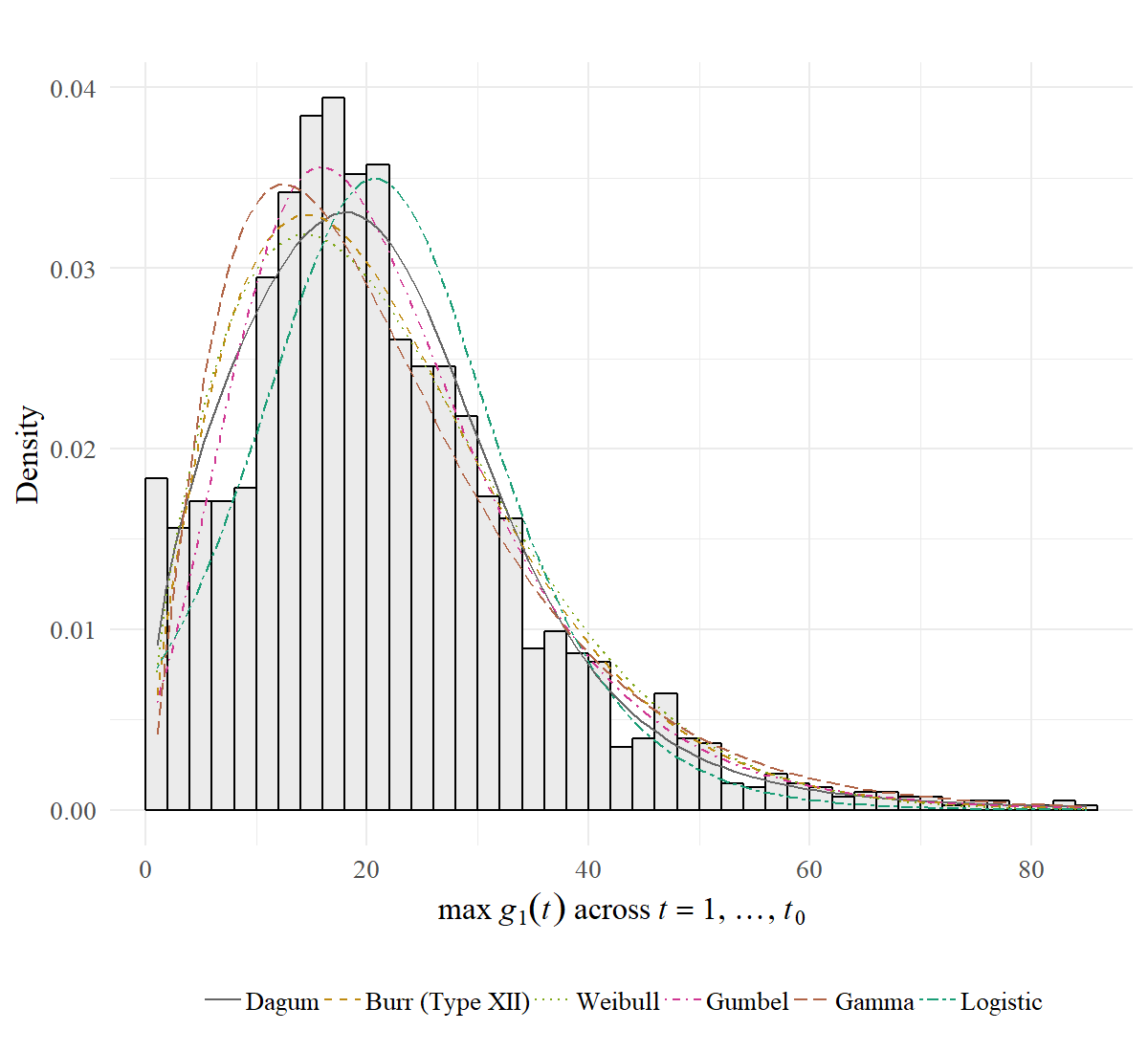}\label{fig:StatFit_b}
\end{subfigure} \vspace{-11pt}
\caption{Candidate statistical distributions that are fit on the maximum of the weighted payments in arrears $\text{max}\,g_1(t)$ observed per account across historical periods $t=1,\dots,t_0$. These maxima are respectively calculated from the $S_2$ sample (delinquents) in \textbf{(a)} and $S_3$ (write-offs) in \textbf{(b)}, with a histogram of maxima given in each case.}\label{fig:StatFit}
\end{figure}
\clearpage
} \clearpage

%TC:endignore

%--------------------------------------------------------%
%	END DOCUMENT
%--------------------------------------------------------%

\end{document}